\documentclass[fleqn,usenatbib]{mnras}

\usepackage[authoryear]{natbib}
\usepackage[T1]{fontenc}
\usepackage{ae,aecompl}
\usepackage[dvipsnames]{xcolor}
\usepackage{psfrag,color}
\usepackage[latin1]{inputenc}
\usepackage{graphicx}	
\usepackage{amsmath}	
\usepackage{amssymb}	
\usepackage[flushleft]{threeparttable}

\newcommand{\dens}{$n_{\rm e}$}
\newcommand{\temp}{$T_{\rm e}$}
\newcommand{\efftemp}{$T_{\rm eff}$}
\newcommand{\lum}{$L_{*}$}
\newcommand{\densH}{$n_{{\rm H}}$}

\newcommand{\ups}{$\upsilon$}
\newcommand{\ome}{$\omega$}

\newcommand{\fnaiii}{[\ion{Na}{iii}]}
\newcommand{\fnaiv}{[\ion{Na}{iv}]}
\newcommand{\fnavi}{[\ion{Na}{vi}]}
\newcommand{\fnavii}{[\ion{Na}{vii}]}
\newcommand{\fcaii}{[\ion{Ca}{ii}]}
\newcommand{\fcaiv}{[\ion{Ca}{iv}]}
\newcommand{\fcav}{[\ion{Ca}{v}]}

\newcommand{\fcaviii}{[\ion{Ca}{viii}]}
\newcommand{\fkiii}{[\ion{K}{iii}]}
\newcommand{\fkiv}{[\ion{K}{iv}]}
\newcommand{\fkv}{[\ion{K}{v}]}
\newcommand{\fkvi}{[\ion{K}{vi}]}
\newcommand{\fkvii}{[\ion{K}{vii}]}

\title[ICFs for Na, K, and Ca in PNe]{Ionization correction factors for sodium, potassium, and calcium in planetary nebulae} 

\author[Amayo et al.]{
A. Amayo,$^{1}$\thanks{E-mail: amedina, gdelgado@astro.unam.mx (AMA, GDI)}
G. Delgado-Inglada,$^{1}$ 
and J. Garc\'ia-Rojas$^{2,3}$
\\
$^{1}$Instituto de Astronom\'ia, Universidad Nacional Aut\'onoma de M\'exico, Ap. 70-264, 04510, CDMX, Mexico\\
$^{2}$Instituto de Astrof\'isica de Canarias, E-38200, La Laguna, Tenerife, Spain\\
$^{3}$Universidad de La Laguna. Depart. de Astrof\'isica, E-38206, La Laguna, Tenerife, Spain\\
}

\date{Accepted XXX. Received YYY; in original form ZZZ}
\pubyear{2019}

\begin{document}

\maketitle

\begin{abstract}
We use a large grid of photoionization models that are representative of observed planetary nebulae (PNe) to derive ionization correction factors (ICFs) for sodium, potassium, and calcium. In addition to the analytical expressions of the ICFs, we provide the range of validity where the ICFs can be safely used and an estimate of the typical uncertainties associated with the ICFs. We improved the previous ICFs for calcium and potassium in the literature and suggest for the first time an ICF for sodium. We tested our ICFs with a sample of 39 PNe with emission lines of some ion of these elements. No obvious trend is found between the derived abundances and the degree of ionization, suggesting that our ICFs do not seem to be introducing an artificial bias in the results. The abundances found in the studied PNe range from $-2.88_{-0.22}^{+0.21}$ to $-2.09\pm0.21$ in $\log$(Na/O), from $-4.20_{-0.45}^{+0.31}$ to $-3.05_{-0.47}^{+0.26}$ in $\log$(K/O), and from $-3.71_{-0.34}^{+0.41}$ to $-1.57_{-0.47}^{+0.33}$ in $\log$(Ca/O). These numbers imply that some of the studied PNe have up to $\sim$65\%, 75\%, or 95\% of their Na, K, and/or Ca atoms condensed into dust grains, respectively. As expected, the highest depletions are found for calcium which is the element with the highest condensation temperature.
\end{abstract}


\section{Introduction}
\label{sec:intro}
Planetary nebulae (PNe) are the final products of many low to intermediate-mass stars, those having initial masses between $\sim$0.8 and 8 $M_\odot$. These ionized nebulae are produced after the ejection of the outer shells of the central star in the thermally pulsing Asymptotic Giant Branch (AGB) phase. If the ultraviolet photons of the star are able to ionize the gas before it is diluted into the interstellar medium (ISM), a PN is formed. 

The abundances of PNe are useful tools to study many aspects of stellar and galactic evolution. Some elements such as helium and nitrogen reflect the nucleosynthesis processes that have occurred during the stellar evolution. Others such as argon and chlorine are not affected by the stellar nucleosynthesis and thus, can be used to trace the composition of the interstellar medium when the progenitors of PNe were formed. 

There are also a few elements that serve us to study dust formation and evolution in PNe. Refractory elements such as iron, nickel, and calcium have gaseous abundances in the ISM that are much lower than a reference value, i.e. solar \citep{Morton1973, Morton1974}. This underabundance is assumed to be the consequence of their atoms being deposited into dust grains. Therefore, by studying the gaseous abundance of refractory elements in PNe we can learn about dust grain composition and their evolution in the ionized medium that forms PNe
\citep[e.g.,][]{Stasinska1999, Phillips2007, delgadoinglada2009, delgadoinglada2014}.

The determination of an element abundance in ionized nebulae relative to hydrogen, X/H, is generally computed by adding up the ionic abundances of all the ions present in the gas. In order to take into account the contribution of unobserved ions (because their emission lines are too weak or because they are emitted in a different spectral range than the observed one), one needs to make use of the ionization correction factors (ICFs). \citet{DI2014, delgadoinglada2016} used a large grid of photoionization models to provide new ICFs for carbon, nitrogen, oxygen, neon, sulphur, chlorine, argon, and nickel to be used in PNe with optical observations.

Here, we use the same grid of models and the same approach to suggest new ICFs for three other elements: sodium, potassium, and calcium.
Due to their condensation temperature, calcium, potassium and sodium (in that order) are most likely to be condensed onto dust grains, leading to bigger depletion factors than other elements such as carbon or lead (\citealt{Savage1996, Field1974}).
Pioneering studies of the gas phase depletions of these elements include works by \citet{Shields1981} and by \citet{Kingdon1995}, who used PNe photoionization models along with observations and obtained calcium depletion factors of $\sim$1.5 and 2.5 dex, respectively.
Other published values for the gaseous abundances of these elements cover the ranges: near solar to 1/2 the solar abundance for sodium, near solar to 1/6 solar for potassium, and 1/2 to 1/40 the solar abundance for calcium (\citealt{Aller1981, Aller1983, Keyes1990, Casassus2000, Hyung2001, Pottasch2007, Bohigas2013}).

The only available ICFs in the literature for these elements are those proposed by \citet{Bohigas2013} for potassium and calcium. However, they are simple expressions based on similarities between the ionization potentials of K$^{+3}$ and Ar$^{+3}$ (60.9 and 59.8 eV, respectively) and Ca$^{+4}$ and Ar$^{+4}$ (84.4 and 75.0 eV, respectively). These simple expressions are not necessarily correct \citep[e.g.,][]{stasinska2002}.  

To test our computed ICFs, we have compiled from the literature a sample of PNe where at least one emission line of some ion of sodium, potassium, and calcium has been detected. Deep, high-resolution optical spectrophotometry making use of instruments attached to 8--10-m type telescopes, has significantly raised the amount of heavy element emission lines in PNe spectra. As an example: in a very deep, high-resolution (R~$\sim40,000$) optical spectrum of the high-excitation PN NGC\,3918, several emission lines of {\fnaiv}, {\fkiv}, {\fkv}, {\fkvi}, and {\fcav} were identified and measured by \citet{GR2015}. In the next decade, the upcoming of giant (30--40-m class) telescopes will increase the number of detected Na, K and Ca emission lines, therefore, the computation of appropriate ICFs using an extensive database of photoionization models will be extremely useful to estimate accurate elemental abundances.

\section{The grid of photoionization models}
\label{sec:models}

We have used a subgrid of models from the Mexican Million Models database (3MdB, \citealt{Morisset2014}).  The subgrid is defined by the models under the ``PNe\_2014\_c13'' and ``PNe\_2016'' references, having a blackbody as the ionization source, solar abundances, no dust, a constant density distribution of the gas, and obtained with the version c13 of the multipurpose photoionization code {\sc cloudy} \citep{Ferland2013}. Since this sample of photoionization models was specially designed to compute the ICFs of several elements in PNe, it is ideal for our purposes. 

The input parameters and the physical and chemical assumptions are wide enough to obtain a grid that is representative of most of the observed PNe. The effective temperature (\efftemp) ranges from 25,000 to 300,000 K, the inner radius (R$_{\rm in}$) from $3\times10^{15}$ to $3\times10^{18}$ cm, the stellar luminosity (\lum) from 200 to 17,800 L$_{\odot}$, and the hydrogen density (\densH) from 30 to 300,000 cm$^{-3}$. The code stops when the fraction of H$^{+}$ decreases below 2\% producing radiation bounded models. \citet{DI2014} computed matter bounded models by cutting the radiation bounded models at 40\%, 60\%, and 80\% of their total gas mass.

The initial grid contains $\sim60,300$ models but we applied the filters described by \citet{DI2014} to remove unrealistic models: 1) We exclude models with hydrogen masses above 1 M$_{\odot}$, 2) we exclude models where the effective temperatures and luminosities fall outside the evolutionary tracks proposed by \citet{Schoenberner1983} and \citet{Bloecker1995} for PN with central star masses between 0.58 to 0.70 M$_{\odot}$, 3) we limit the H$\beta$ surface brightness to the range $10^{-13} - 10^{-11}$ erg s$^{-1}$cm$^{-2}$arcsec$^{-2}$, 4) we also exclude models with large outer radius and high electron density as well as the opposite case. \footnote{The models corresponding to these criteria are obtained with the com6 $=1$ condition in 3MdB}. In addition, we only consider here the radiation bounded models and the matter bounded models constructed by cutting the radiation bounded models at 80\% of their total gas mass disconsidering the ones in the initial grid whose cut is more severe. The reason is that we identified that the heavily cut models were those which led to an unphysical relation between N/O and He/H (upper panel of Fig. 5 in \citealt{delgadoinglada2015}). This leads us to a final sample of 3,916 photoionization models from which we compute the ICFs. 

\begin{figure}
\includegraphics[trim=0.4cm 0.6cm 0cm 0cm, clip=true, width=1.04\columnwidth]{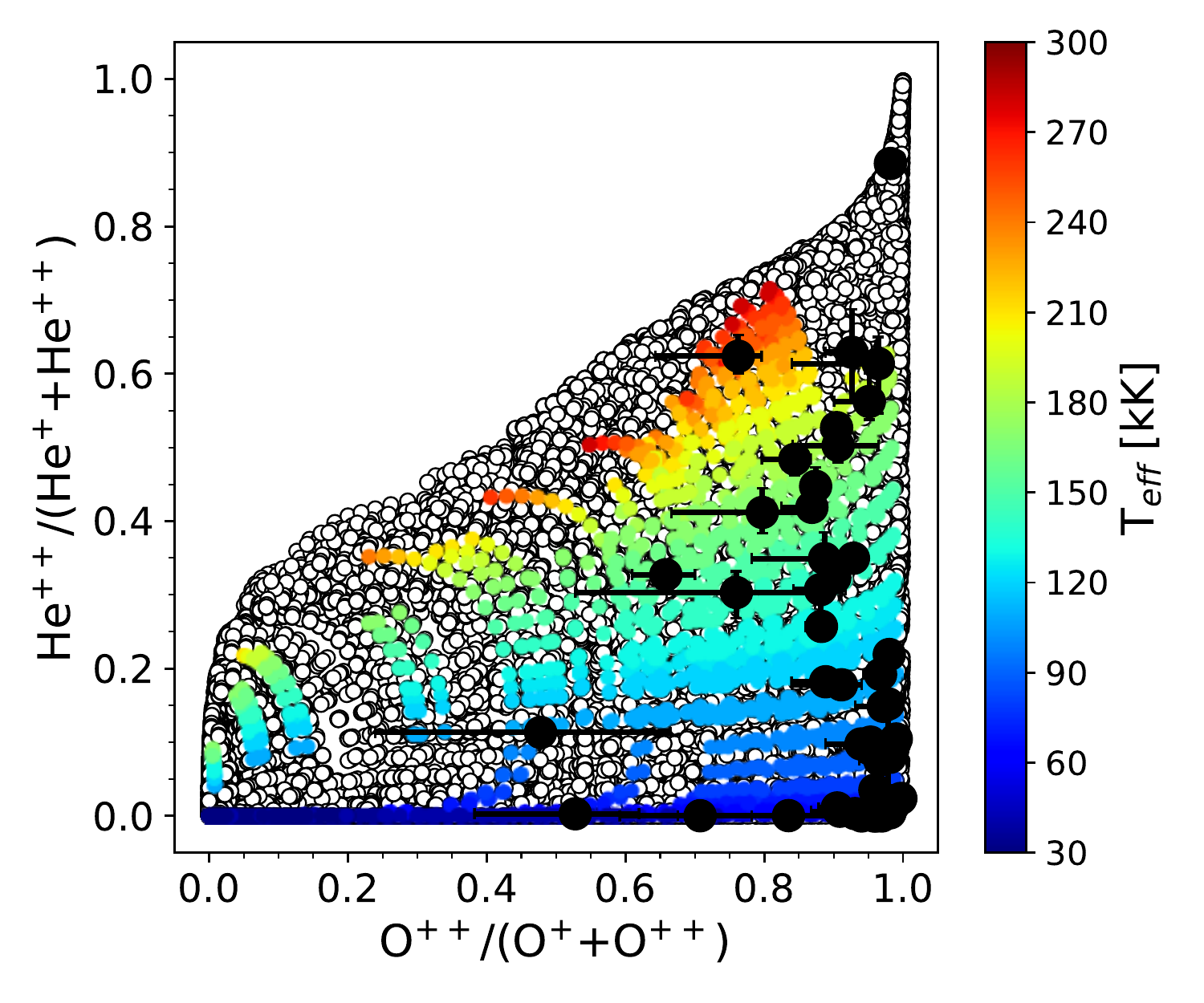} 
\caption{Values of He$^{++}$/(He$^{+}$+He$^{++}$) as a function of O$^{++}$/(O$^{+}$+O$^{++}$) for the initial (empty circles) and final (coloured circles) grid of photoionization models. Black points represent the observed objects used in this work, detailed in section~\ref{sec:ObsSample}. The colorbar located on the side runs from low to high values of the effective temperature of the central star.}
\label{fig:gihe_gio}
\end{figure}

Figure~\ref{fig:gihe_gio} shows the values of He$^{++}$/(He$^{+}$+He$^{++}$) as a function of O$^{++}$/(O$^{+}$+O$^{++}$) for the initial and final samples of photoionization models along with the observational sample used here (black circles). It can be seen that the final grid of models covers a wide range in degree of ionization. In addition, the grid covers most of the observational sample used here with the exception of NGC\,2022 (taken from the observations of \citealt{Tsamis2003}). Therefore, we consider that these models are representative of most of the observed PNe and hence, are adequate to derive ICFs.

\section{Methods and notations}
\label{sec:methods}

 The total abundance of one particular element with respect to hydrogen abundance, $X$/H, is computed as the sum of all the ionic abundances ($X^{+i}$/H$^+$) of the ions present in a nebula. The ionic abundances are calculated as:
\begin{equation}
\frac{X^{+i}}{\rm {H}^+} = \frac{I_\lambda}{I(\rm {H}\beta)} \frac{\epsilon(\rm {H}\beta)}{\epsilon(\lambda)}
\end{equation}
where $\epsilon$($\rm {H}\beta$) and $\epsilon$($\lambda$) correspond to the emissivities of H$\beta$ and one line emitted by the ion involved, and $I$($\rm {H}\beta$) and $I_\lambda$ are the line intensities of H$\beta$ and the measured line. 

In general, not all the ions present in ionized nebulae can be observed, either because their emission lines are emitted in a different wavelength range than the observed one or because the lines are too weak to be measured. In such cases, it is necessary to use an ionization correction factor (ICF) that take into account the contribution of unobserved ions to the total abundance. 

The quantities and notations used here are similar to those presented in Section 4.1 of \citet{DI2014}. We will use one of the cases we studied here as an example to explain the notation adopted. If the only observed ion of sodium is Na$^{++}$ the total abundance can be expressed as:
\begin{equation}\label{eq:abna_1}
\frac{{\rm Na}}{{\rm H}} = \frac{{\rm Na^{++}}}{{\rm H^{+}}} \times {\rm ICF}({\rm Na^{++}})
\end{equation}
and hence the ICF is given by:
\begin{equation}\label{eq:icfna_1}
{\rm ICF}({\rm Na^{++}}) = \frac{x({\rm H^{+}})}{x({\rm Na^{++}})}, 
\end{equation}
where $x$ corresponds to the relative ionic fractions weighted by the electron density \dens, which are one of the outputs of the photoionization code {\sc cloudy} available in 3MdB. This expression is called ICF$_{\rm m}$. We use the photoionization models to look for correlations between this quantity and other ionic fractions that are easily obtained from observations. Following the work by \citet{DI2014}, we have tried to use these two ionic fractions:
\begin{equation}\label{eq:omega}
\omega = \frac{{\rm O^{++}}}{({\rm O^{+}+\rm O^{++})}}
\end{equation}
and
\begin{equation}\label{eq:upsilon}
\upsilon = \frac{{\rm He^{++}}}{({\rm He^{+}+\rm He^{++})}}.
\end{equation}
We perform a fit to the best correlation and provide the analytical expressions of the ICFs, ICF$_{\rm f}$. In some cases, other options, i.e., fits depending on other parameters different from $\omega$ and $\upsilon$, are suggested. In addition, we also illustrate the magnitude of the uncertainties associated with the proposed ICFs. The uncertainties in dex in the total abundances are given by $\Delta_{{\rm ICF}}$ = $\log$ICF$_{\rm f}$ - $\log$ICF$_{\rm m}$. 

\section{Ionization correction factors}\label{sec:ICF}
\subsection{Sodium}
The sodium lines that have been more frequently observed in PNe are: [\ion{Na}{iii}] $\lambda$7.32 $\mu$m and [\ion{Na}{iv}] $\lambda$3242, $\lambda$3362 \citep{Hyung1994, Hyung1995, Hyung2001, Tsamis2003, Pottasch2003b, Pottasch2007, Pottasch2009, GR2015} but other lines have also been detected (see Table~\ref{tab:1}). According to our grid of photoionization models, the ions that contribute the most to the total Na abundance are Na$^{+}$, Na$^{++}$, Na$^{+3}$, and Na$^{+4}$ (with ionization potentials, I.P., 47.3, 71.6, 98.9, and 138.4 eV, respectively). We do not expect any significant contribution of Na$^{+5}$ (I.P. = 172.2 eV) and Na$^{+6}$ (I.P. = 208.5 eV).

For this element we have derived three correction schemes depending on the observed lines: 1) to be used when [\ion{Na}{iii}] lines are observed, 2) to be used when [\ion{Na}{iv}] lines are observed, and 3) to be used when [\ion{Na}{iii}] and [\ion{Na}{iv}] lines are observed. Note that scheme 3) may imply to use infrared and optical observations together.

\citet{BeintemaPott1999} reported the detection of the infrared lines: [\ion{Na}{iv}] $\lambda$9.0 and 21.3 $\mu$m, [\ion{Na}{vi}] $\lambda$8.6 and 14.39 $\mu$m, and [\ion{Na}{vii}] $\lambda$4.7 $\mu$m in NGC~6302. Since these lines have been detected only by these authors and only in this PN, we do not provide an ICF to be used in this case.

\subsubsection{\rm ICF based on Na$^{++}$}
\begin{figure}
\includegraphics[trim=0.2cm 1.2cm 1.cm 0.2cm, clip, width=\columnwidth]{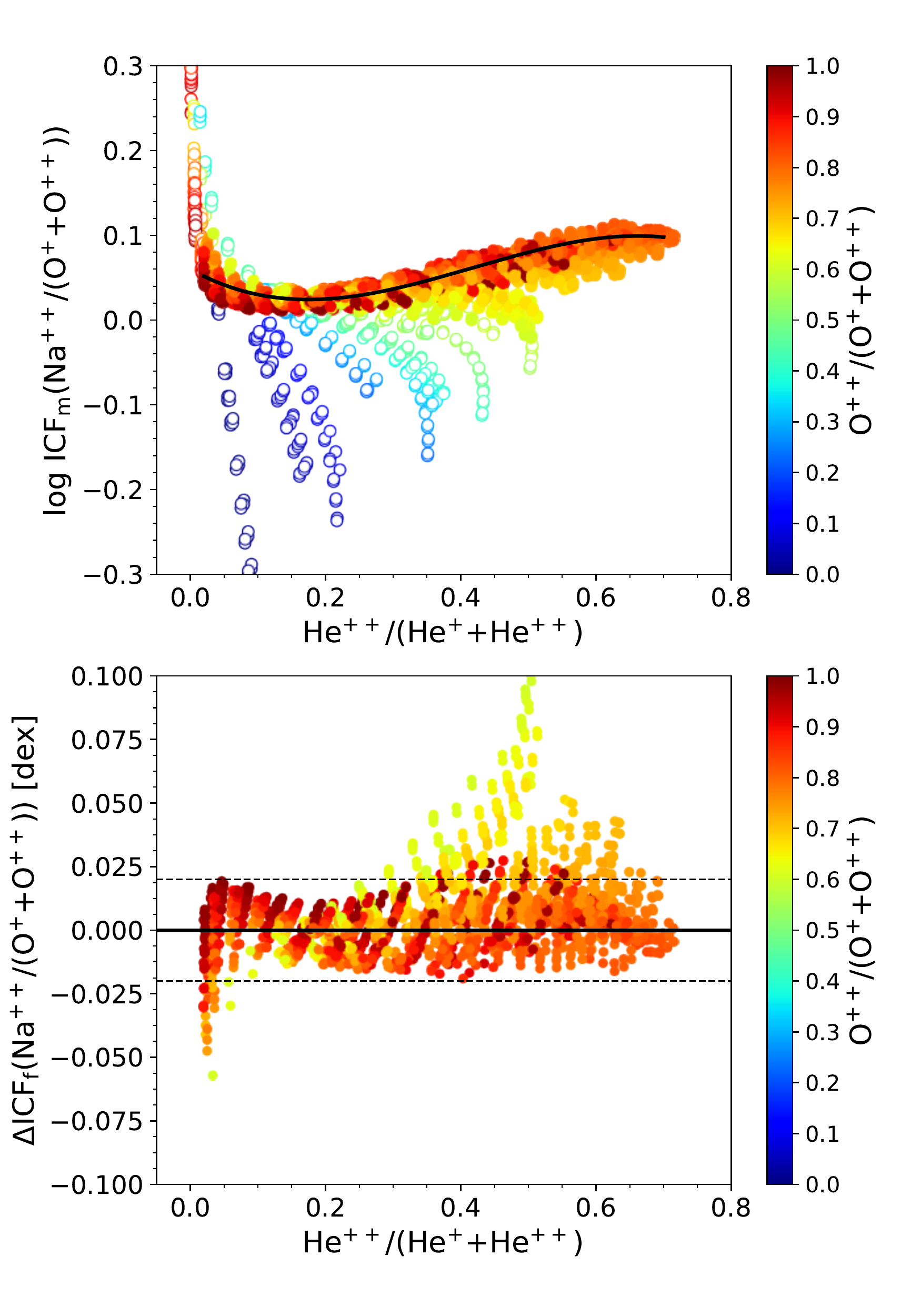}
\caption{{\it Upper panel:} Values of ICF$_{\rm m}$(Na$^{++}$/(O$^{+}$+O$^{++}$)) as a function of He$^{++}$/(He$^{+}$+He$^{++}$) for our photoionization models. The line represents ICF$_{\rm f}$(Na$^{++}$/(O$^{+}$+O$^{++}$)) from equation~\ref{eq:na_1} derived from models with \ome~$\geq0.6$ and \ups~$\geq 0.02$ (represented with filled circles). {\it Lower panel:} Values of $\Delta$ICF$_{\rm f}$(Na$^{++}$/(O$^{+}$+O$^{++}$)) as a function of He$^{++}$/(He$^{+}$+He$^{++}$) for the photoionization models with \ome~$\geq0.6$ and \ups~$\geq0.02$. The solid black line represents where ICF$_{\rm m}$(Na$^{++}$/(O$^{+}$+O$^{++}$)) = ICF$_{\rm f}$(Na$^{++}$/(O$^{+}$+O$^{++}$)). The dashed lines represent the uncertainty range adopted for this ICF. The colorbar located on the right side of both panels runs from low to high values of O$^{++}$/(O$^{+}$+O$^{++}$).}
\label{fig:na_1}
\end{figure}

We suggest two ICFs for this case. The first one:
\begin{equation}\label{eq:na_1}
\log {\rm ICF}_{\rm{f}}\left(\frac{\rm{Na}^{++}}{\rm{O}^{+}+\rm{O}^{++}}\right) = 0.06 - 0.45\upsilon + 1.63\upsilon^{2} -1.30\upsilon^{3}
\end{equation}
can be used when \ome~$\geq 0.6$ and \ups~$\geq 0.02$. Figure~\ref{fig:na_1} shows the values of ICF$_{\rm m}$(Na$^{++}$/(O$^{+}$+O$^{++}$)) as a function of He$^{++}$/(He$^{+}$+He$^{++}$). The ICF from Equation~\ref{eq:na_1} (solid line) is shown in the upper panel together with the photoionization models (circles). Models with \ome~$\geq0.6$ and \ups~$\geq 0.02$ are represented with filled circles whereas the others are shown with empty symbols. As can been seen, the models with \ome~$<0.6$ and \ups~$<0.02$ are not fitted by equation.~\ref{eq:na_1} and thus, we recommend not using this ICF outside the proposed range.

The uncertainties in $\log$(Na/O) associated with this ICF are presented in the lower panel of Figure~\ref{fig:na_1} only for the models within the range of validity: \ome~$\geq 0.6$ and \ups~$\geq 0.02$. For most of the models they are $\pm0.02$ dex so this can be taken as a typical uncertainty in $\log$(Na/O) associated with the use of this ICF.  

\begin{figure}
\includegraphics[trim=0.4cm 1.2cm 0.8cm 0.2cm, clip, width=\columnwidth]{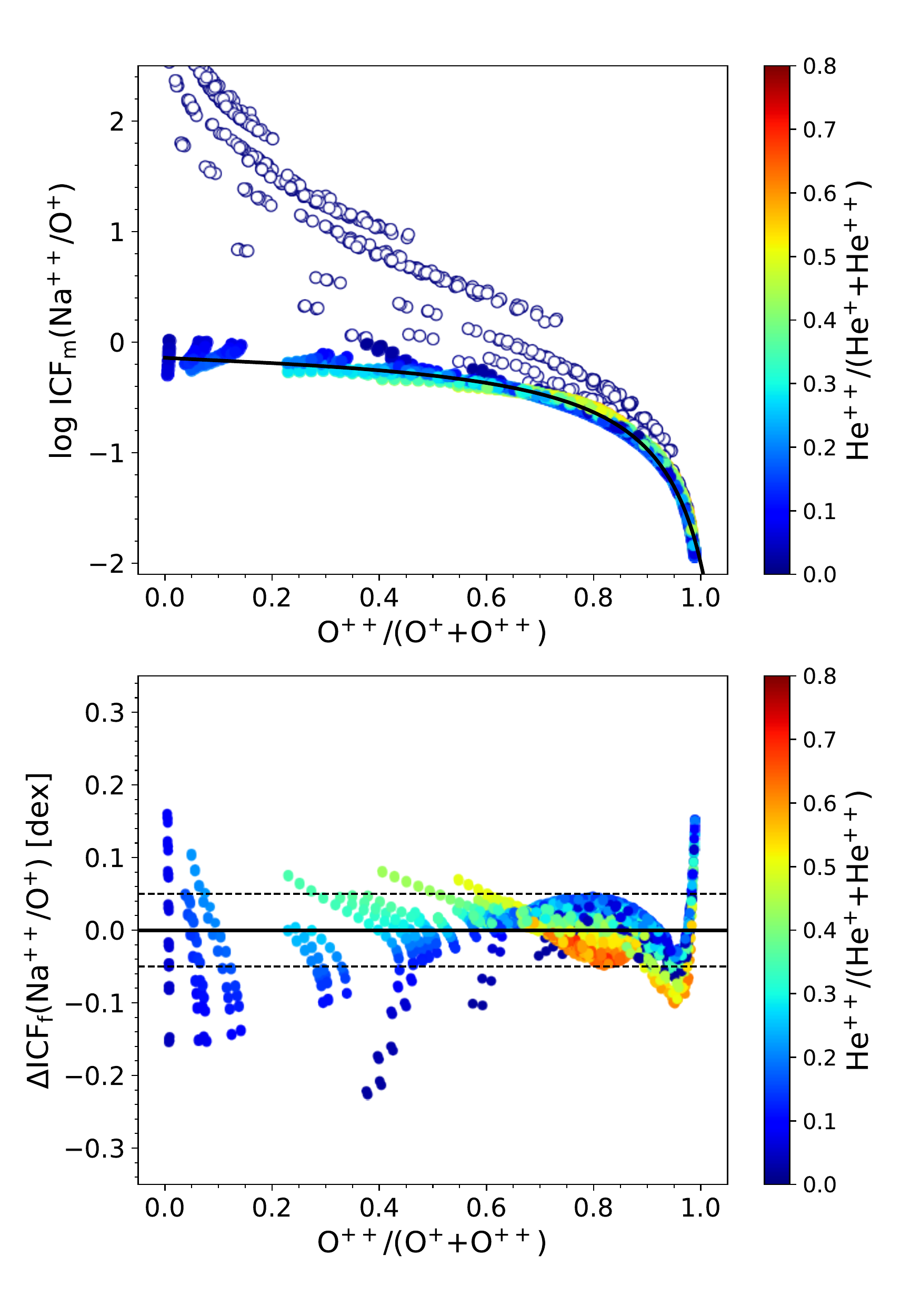}
\caption{{\it Upper panel:} Values of ICF$_{\rm m}$(Na$^{++}$/O$^{+}$) as a function of O$^{++}$/(O$^{+}$+O$^{++}$) for our photoionization models. The line represents ICF$_{\rm f}$(Na$^{++}$/O$^{+}$) from equation~\ref{eq:na_2} derived from models with and \ups~$\geq 0.02$ (represented with filled circles). {\it Lower panel:} Values of $\Delta$ICF$_{\rm f}$(Na$^{++}$/O$^{+}$) as a function of O$^{++}$/(O$^{+}$+O$^{++}$) for the photoionization models with \ups~$\geq 0.02$. The solid black line represents where ICF$_{\rm m}$(Na$^{++}$/O$^{+}$) = ICF$_{\rm f}$(Na$^{++}$/O$^{+}$). The dashed lines represent the uncertainty range adopted for this ICF. The colorbar located on the right side of both panels runs from low to high values of He$^{++}$/(He$^{+}$+He$^{++}$).}
\label{fig:na_2}
\end{figure}

We also derived an alternative ICF that can be used in a larger number of PNe, those with \ups~$\geq 0.02$, but that has somewhat larger uncertainties associated. The upper panel of Figure~\ref{fig:na_2} shows the values of ICF$_{\rm m}$(Na$^{++}$/O$^{+}$) as a function of O$^{++}$/(O$^{+}$+O$^{++}$). The models with \ups~$\geq 0.02$ are represented with filled circles whereas the others are shown with empty symbols. The analytical expression we obtained (solid line) is given by:
\begin{equation}\label{eq:na_2}
\log {\rm ICF}_{\rm{f}}\left(\frac{\rm{Na}^{++}}{\rm{O}^{+}}\right) = \frac{0.15}{\omega^{0.80} - 1.08}.
\end{equation}
The uncertainties in $\log$(Na/O) associated with this ICF are presented in the lower panel only for the models with \ups~$\geq 0.02$. It can be seen that $\pm0.05$ dex can be taken as a typical uncertainty associated with this ICF.  

\subsubsection{\rm ICF based on Na$^{+3}$}

\begin{figure}
\includegraphics[trim=0.5cm 1.2cm 1.cm 0.2cm, clip, width=\columnwidth]{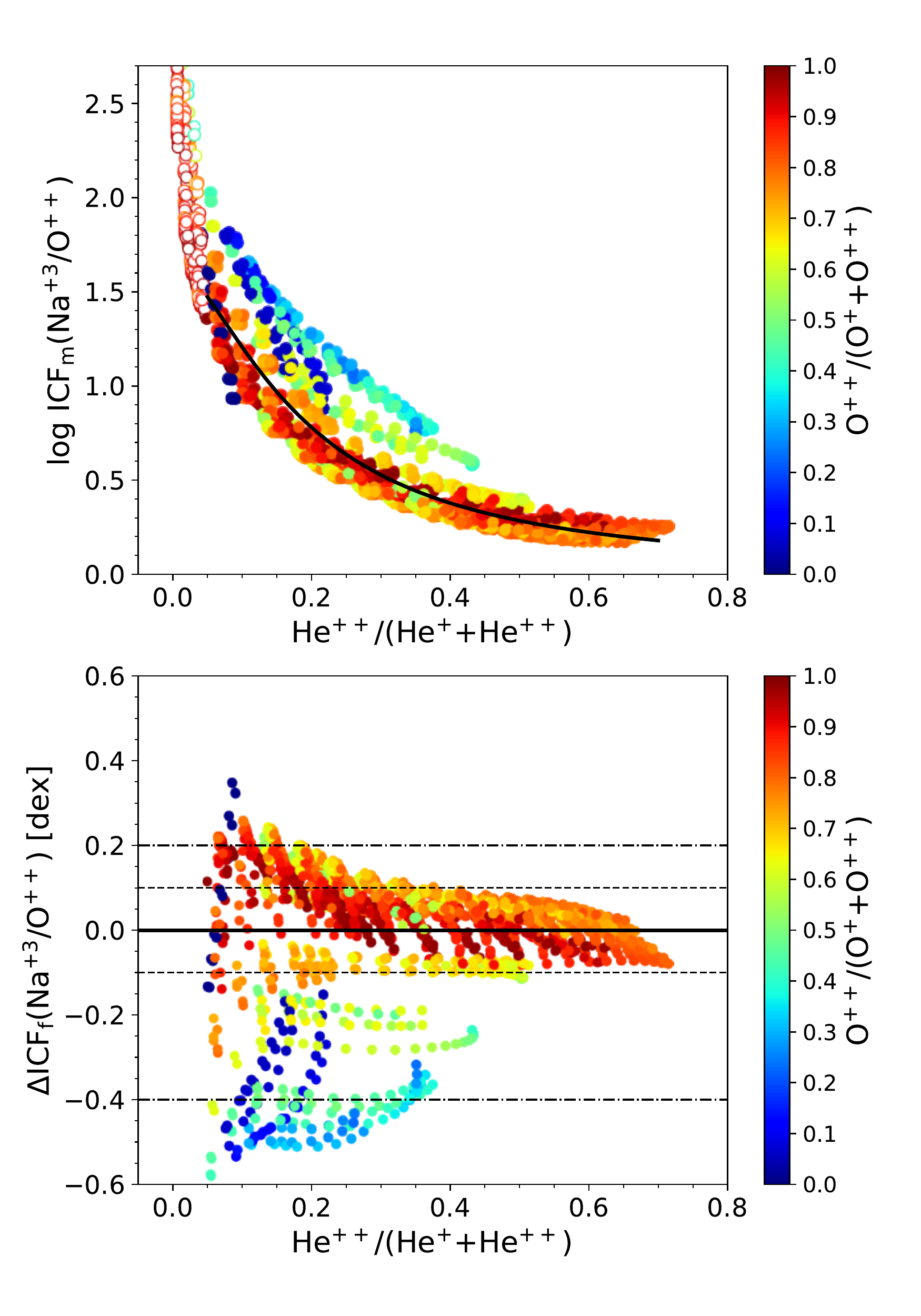}
\caption{{\it Upper panel:} Values of ICF$_{\rm m}$(Na$^{+3}$/O$^{++}$) as a function of He$^{++}$/(He$^{+}$+He$^{++}$) for our photoionization models. The line represents ICF$_{\rm f}$(Na$^{+3}$/O$^{++}$) from equation~\ref{eq:na_3} derived using only the models with \ups~$\geq0.05$ (represented with filled circles). {\it Lower panel:} Values of $\Delta$ICF$_{\rm f}$(Na$^{+3}$/O$^{++}$) as a function of He$^{++}$/(He$^{+}$+He$^{++}$) for the photoionization models with \ups~$\geq0.05$. The solid black line represents where ICF$_{\rm m}$(Na$^{+3}$/O$^{++}$) = ICF$_{\rm f}$(Na$^{+3}$/O$^{++}$). The dashed lines represent the uncertainty range recommended for this ICF when \ups~$>0.5$. The colorbar located on the right side of both panels runs from low to high values of O$^{++}$/(O$^{+}$+O$^{++}$).}
\label{fig:na_3}
\end{figure}

The best correlation we found between the Na$^{+3}$ ionic fraction and  \ups\  or \ome\  is showed in the upper panel of Figure~\ref{fig:na_3}. The fit (black solid line) was obtained taking into account only the models with \ups~$\geq 0.05$:
\begin{equation}\label{eq:na_3}
\log {\rm ICF}_{\rm{f}}\left(\frac{{\rm Na}^{+3}}{{\rm O}^{++}}\right) = \frac{0.11}{0.07 + 1.59\upsilon}.
\end{equation}
The uncertainties in $\log$(Na/O) associated with this ICF (lower panel of Figure~\ref{fig:na_3}) are around $^{+0.2}_{-0.4}$ dex (dash-dotted lines shown in Figure~\ref{fig:na_3}). For the models with \ups~$>0.5$ the uncertainties are $\pm0.1$ dex (dashed lines shown in Figure~\ref{fig:na_3}). 

\subsubsection{\rm ICF based on Na$^{++}$ and Na$^{+3}$}

\begin{figure}
\includegraphics[trim=0cm 1.2cm 1.cm 0.2cm, clip, width=\columnwidth]{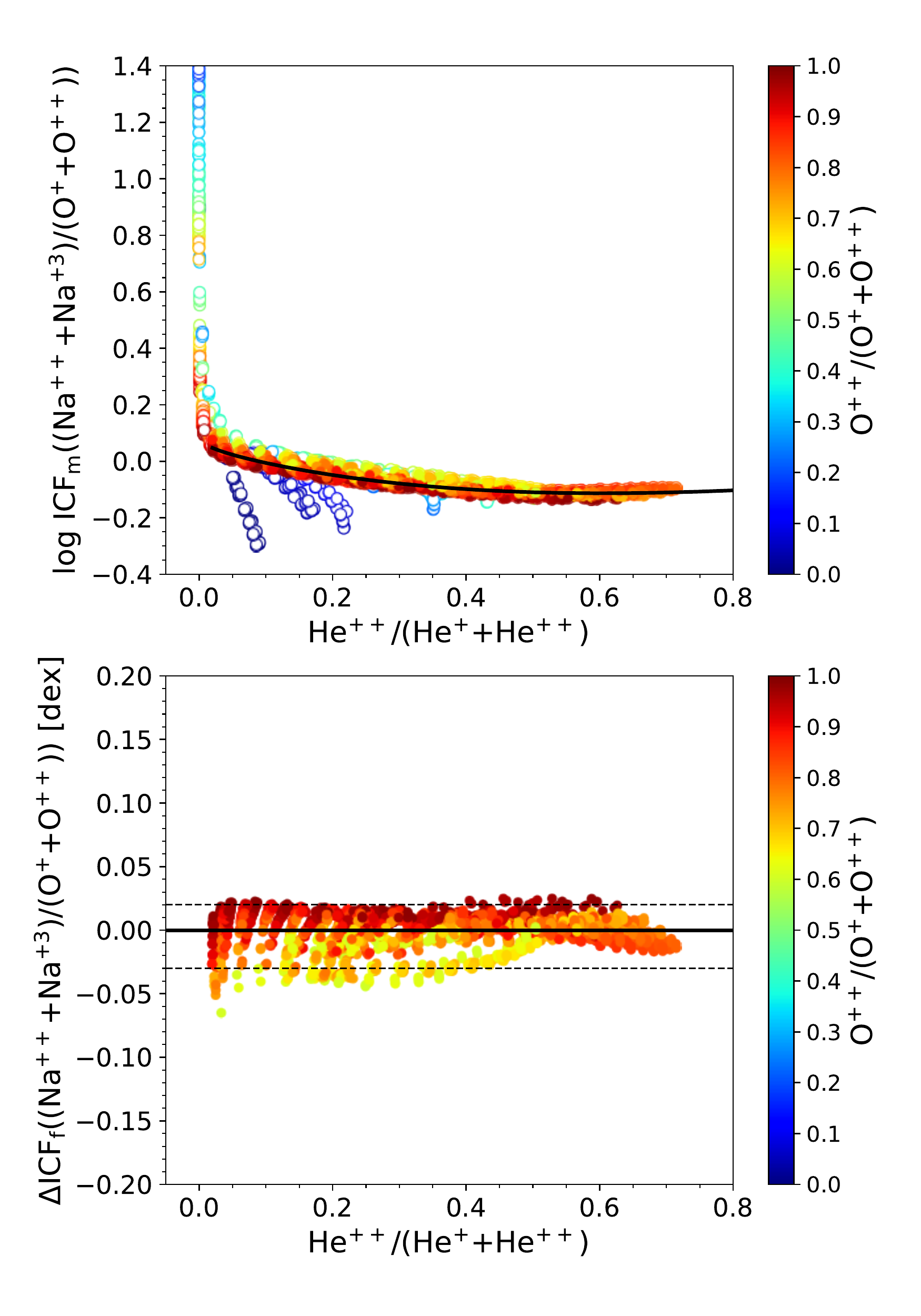}
\caption{{\it Upper panel:} Values of  ICF$_{\rm m}$((Na$^{++}$+Na$^{+3}$)/(O$^{+}$+O$^{++}$)) as a function of He$^{++}$/(He$^{+}$+He$^{++}$) for our photoionization models. The line represents ICF$_{\rm f}$((Na$^{++}$+Na$^{+3}$)/(O$^{+}$+O$^{++}$)) from equation~\ref{eq:na_4} derived using only the models with \ome~$\geq 0.6$ and \ups~$\geq 0.02$ (represented with filled circles). {\it Lower panel:} Values of $\Delta$ICF$_{\rm f}$((Na$^{++}$+Na$^{+3}$)/(O$^{+}$+O$^{++}$)) as a function of He$^{++}$/(He$^{+}$+He$^{++}$) for the photoionization models with \ome~$\geq 0.6$ and \ups~$\geq 0.02$. The solid black line represents where ICF$_{\rm m}$((Na$^{++}$+Na$^{+3}$)/(O$^{+}$+O$^{++}$)) = ICF$_{\rm f}$((Na$^{++}$+Na$^{+3}$)/(O$^{+}$+O$^{++}$)). The dashed lines represent the uncertainty range adopted for this ICF. The colorbar located on the right side of both panels runs from low to high values of O$^{++}$/(O$^{+}$+O$^{++}$).}
\label{fig:na_4}
\end{figure}

In this case we also suggest two ICFs. The first one is more restricted and it was obtained by fitting the correlation between ICF$_{\rm m}$((Na$^{++}$+Na$^{+3}$)/(O$^{+}$+O$^{++}$)) and He$^{++}$/(He$^{+}$+He$^{++}$):
\begin{equation}\label{eq:na_4}
\log {\rm ICF}_{\rm{f}}\left(\frac{\rm{Na}^{++}+\rm{Na}^{+3}}{\rm{O}^{+}+\rm{O}^{++}}\right) = 0.09 - 0.17\times\upsilon^{0.36-1.19\upsilon}.
\end{equation}
It can be used only if \ome~$\geq 0.6$ and \ups~$\geq 0.02$. The upper panel of Figure~\ref{fig:na_4} shows the values of ICF$_{\rm{m}}$ as a function of \ups\ for all the photoionization models. The fit is represented with a solid line. As can been seen, the models with \ome~$<0.6$ and \ups~$<0.02$ (empty circles) are not fitted by the previous expression and thus, we do not recommend to use this ICF in very low excitation PNe and/or relatively low ionized PNe. The uncertainties associated with the use of this ICF are around $^{+0.02}_{-0.03}$ dex.

\begin{figure}
\includegraphics[trim=0.5cm 1.2cm 1.cm 0.2cm, width=\columnwidth]{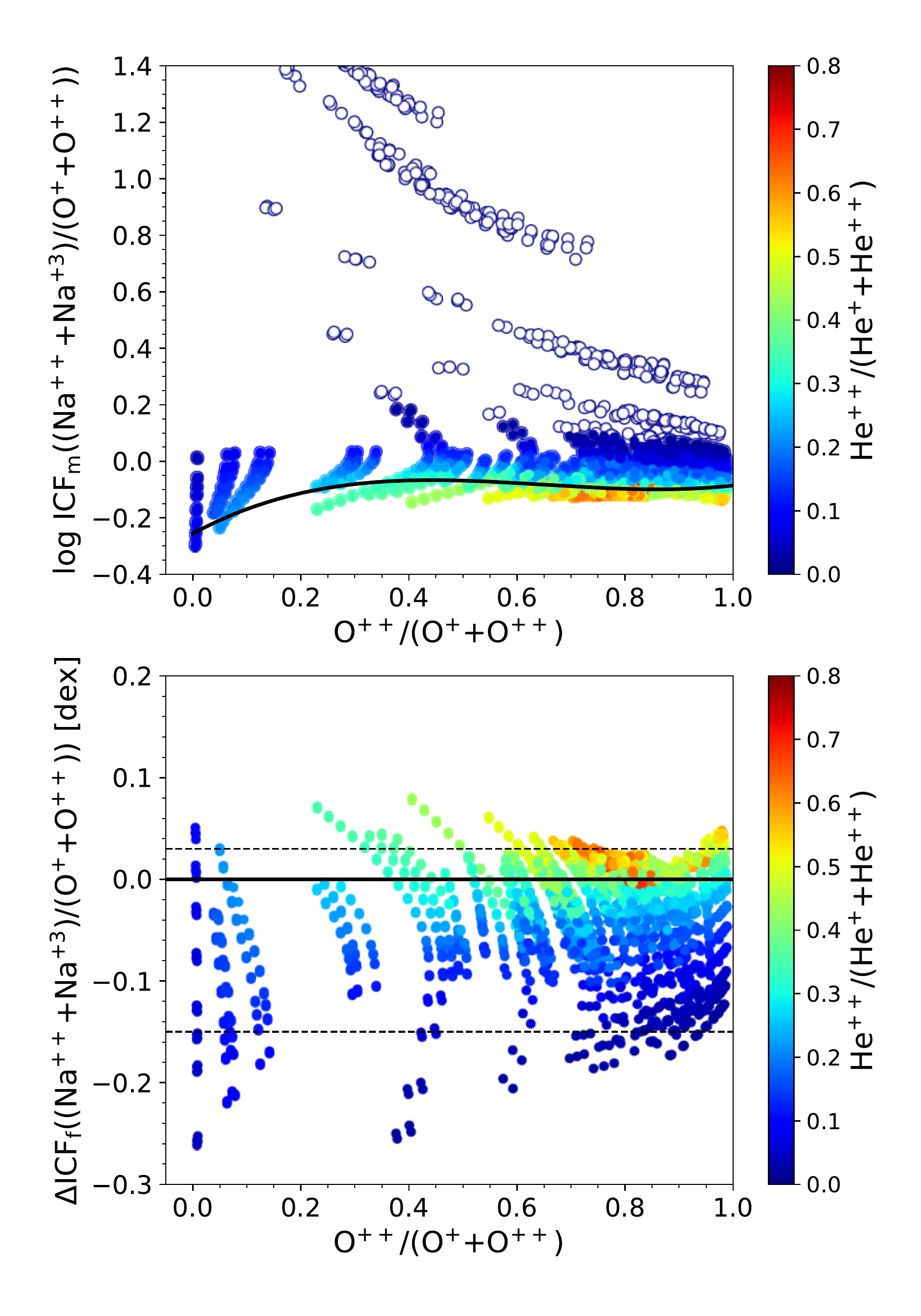}
\caption{{\it Upper panel:} Values of  ICF$_{\rm m}$((Na$^{++}$+Na$^{+3}$)/(O$^{+}$+O$^{++}$)) as a function of O$^{++}$/(O$^{+}$+O$^{++}$) for our photoionization models. The line represents ICF$_{\rm f}$((Na$^{++}$+Na$^{+3}$)/(O$^{+}$+O$^{++}$)) from equation~\ref{eq:na_5} derived using only the models with \ups~$\geq 0.02$ (represented with filled circles). {\it Lower panel:} Values of $\Delta$ICF$_{\rm f}$((Na$^{++}$+Na$^{+3}$)/(O$^{+}$+O$^{++}$)) as a function of  O$^{++}$/(O$^{+}$+O$^{++}$) for the photoionization models with \ups~$\geq 0.02$. The solid black line represents where ICF$_{\rm m}$((Na$^{++}$+Na$^{+3}$)/(O$^{+}$+O$^{++}$)) = ICF$_{\rm f}$((Na$^{++}$+Na$^{+3}$)/(O$^{+}$+O$^{++}$)). The dashed lines represent the uncertainty range adopted for this ICF. The colorbar located on the right side of both panels runs from low to high values of He$^{++}$/(He$^{+}$+He$^{++}$).}
\label{fig:na_5}
\end{figure}

The second proposed ICF was obtained by fitting the correlation between ICF$_{\rm m}$((Na$^{++}$+Na$^{+3}$)/(O$^{+}$+O$^{++}$)) and O$^{++}$/(O$^{+}$+O$^{++}$) and it can be used in all the PNe with \ups~$\geq 0.02$:
\begin{equation}\label{eq:na_5}
\log {\rm ICF}_{\rm{f}}\left(\frac{\rm{Na}^{++}+\rm{Na}^{+3}}{\rm{O}^{+}+\rm{O}^{++}}\right) = -0.26 + 1.03\omega - 1.77\omega^{2} + 0.91\omega^{3}.
\end{equation}
The upper panel of Figure~\ref{fig:na_5} shows the values of ICF$_{\rm{m}}$ as a function of \ome\ for all the photoionization models. The fit is represented with a solid line. The uncertainties associated with the use of this ICF are $^{+0.03}_{-0.15}$ dex.

\subsection{Potassium}

The emission lines more frequently reported in the literature are the following: [\ion{K}{iv}] $\lambda4511$, $\lambda6102$, $\lambda6795$, $\lambda6$ $\mu$m; [\ion{K}{v}] $\lambda4123$, $\lambda4163$; and [\ion{K}{vi}] $\lambda5602$, $\lambda6229$ \citep{Aller1978, Hyung1994, Hyung1995, Aller1999, Kwitter2001, Hyung2001, Pottasch2003a, Tsamis2003, Wesson2004, Pottasch2009, GR2009, GR2012, Bohigas2013, GR2015, GR2018, Wesson2018}. 

After examining all the possibilities, we decided to compute only one ICF based on K$^{+3}$ ionic abundances since using also K$^{+4}$ and K$^{+5}$ ions did not significantly affect the analytical fit that could be provided. This is a consequence of the high ionization potentials of K$^{+4}$ and K$^{+5}$ (82.7 and 100 eV, respectively) that make these ions modest contributors to the total abundance of potassium. These two ions contribute up to 25\% of the potassium abundance in our models with \ome~$> 0.5$. In fact [\ion{K}{v}] and [\ion{K}{vi}] lines have been reported only in one PN.

\citet{BeintemaPott1999} reported the following infrared lines in NGC~6302: [\ion{K}{iii}] $\lambda4.62$ $\mu$m; [\ion{K}{iv}] $\lambda5.98$ $\mu$m, $\lambda15.38$ $\mu$m; [\ion{K}{vi}] $\lambda5.58$ $\mu$m, $\lambda8.83$ $\mu$m; and [\ion{K}{vii}] $\lambda3.19$ $\mu$m. Since this is the only reference where these lines have been detected, we do not provide an ICF for this particular case. However, the ICF proposed here can be applied in such cases.  

\begin{figure}
\includegraphics[trim=0.5cm 1.2cm 1.cm 0.2cm, width=\columnwidth]{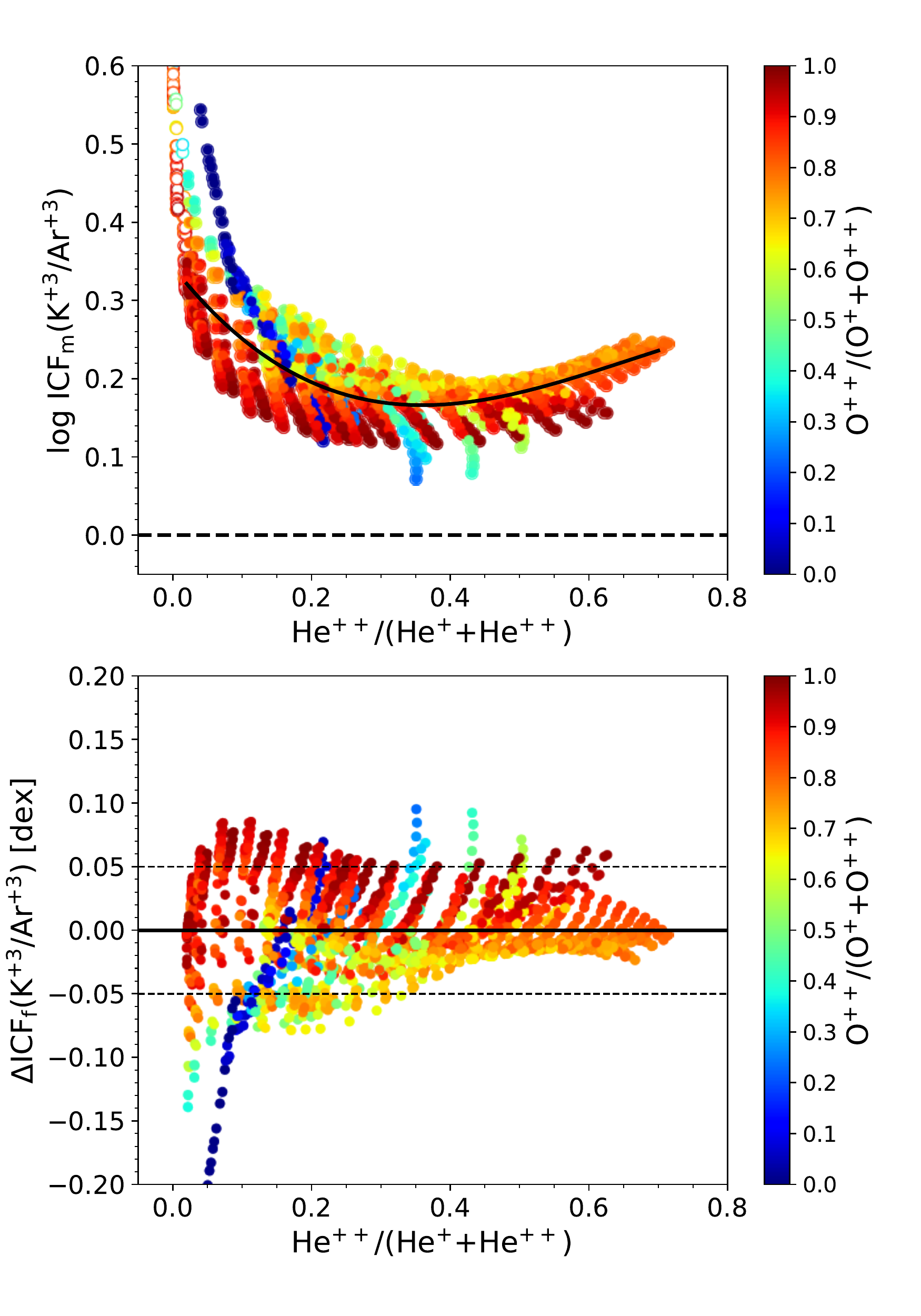}
\caption{{\it Upper panel:} Values of ICF$_{\rm m}$(K$^{+3}$/Ar$^{+3}$) as a function of He$^{++}$/(He$^{+}$+He$^{++}$) for our photoionization models. The dashed line represents the ICF proposed by \citet{Bohigas2013}. The solid line represents ICF$_{\rm f}$(K$^{+3}$/Ar$^{+3}$) from equation~\ref{eq:k_1} derived using only the models with \ups~$\geq 0.02$ (represented with filled circles). {\it Lower panel:} Values of $\Delta$ICF$_{\rm f}$(K$^{+3}$/Ar$^{+3}$) as a function of He$^{++}$/(He$^{+}$+He$^{++}$). The solid black line represents where ICF$_{\rm m}$(K$^{+3}$/Ar$^{+3}$) = ICF$_{\rm f}$(K$^{+3}$/Ar$^{+3}$). The dashed lines represent the uncertainty range adopted for this ICF. The colorbar located on the right side of both panels runs from low to high values of O$^{++}$/(O$^{+}$+O$^{++}$).}\label{fig:k_1}
\end{figure}

We suggest an improvement of the ICF proposed by \citet{Bohigas2013} since we did not find a better ICF for K/H or K/O based on O$^{++}$/(O$^{+}$+O$^{++}$) and He$^{++}$/(He$^{+}$+He$^{++}$). These authors proposed to use the expression K/Ar = K$^{+3}$/Ar$^{+3}$ that it is based on the similarities between the I.P. of K$^{+3}$ (60.9 eV) and Ar$^{+3}$ (59.8 eV). Figure~\ref{fig:k_1} shows ICF$_{\rm m}$(K$^{+3}$/Ar$^{+3}$) as a function of He$^{++}$/(He$^{+}$+He$^{++}$) for our photoionization models. The dashed line represents the ICF suggested by \citet{Bohigas2013}. It is clear from the figure that this correction scheme underestimates the K abundances in all the cases by more than 0.1 dex.

From our grid of photoionization models we obtain the following ICF (solid line in Fig.~\ref{fig:k_1}): 
\begin{equation}
\log{\rm ICF}_{\rm f}\left(\frac{{\rm K}^{+3}}{{\rm Ar}^{+3}}\right) = 0.343-1.124\upsilon+2.15\upsilon^2-1.09\upsilon^3.
\label{eq:k_1}
\end{equation}
This expression can be used when \ups~$\geq 0.02$ (filled circles in upper panel of Fig.~\ref{fig:k_1}). Lower panel of Figure~\ref{fig:k_1} shows that the uncertainties associated with this ICF are, in general, $\pm0.05$ dex. It should be noted that since this ICF is based on the ionic fraction of Ar$^{+3}$, a reliable determination of the total Ar abundance is required. 

\subsection{Calcium}
\label{sec:Ca}
The Ca lines that are more frequently observed in PNe are [\ion{Ca}{ii}] $\lambda7291$, $\lambda7324$ and [\ion{Ca}{v}] $\lambda5309$, $\lambda6086$ \citep{Aller1978, Hyung1994, Hyung1995, Aller1999, Kwitter2001, Pottasch2003a, GR2012, Bohigas2013, GR2015, GR2018}. Since the I.P. of Ca$^{+}$ is 11.9 eV, this ion is mostly located in the photodissociation region and since this region is not included in our models, they are not adequate to compute one ICF for calcium based on Ca$^{+}$ abundance. Therefore, we only compute one ICF to be used when only [\ion{Ca}{v}] lines are detected.

\citet{BeintemaPott1999} identified other Ca lines in NGC~6302: [\ion{Ca}{iv}] $\lambda3.21$ $\mu$m, [\ion{Ca}{v}] $\lambda4.16$ $\mu$m, $\lambda11.49$ $\mu$m, [\ion{Ca}{viii}] $\lambda6.15$ $\mu$m. We do not consider these ions because this is the only object where these lines have been identified.

\subsubsection{\rm Ca$^{+4}$}

\begin{figure}
\includegraphics[trim=0.2cm 1.2cm 1.cm 0.2cm, clip, width=\columnwidth]{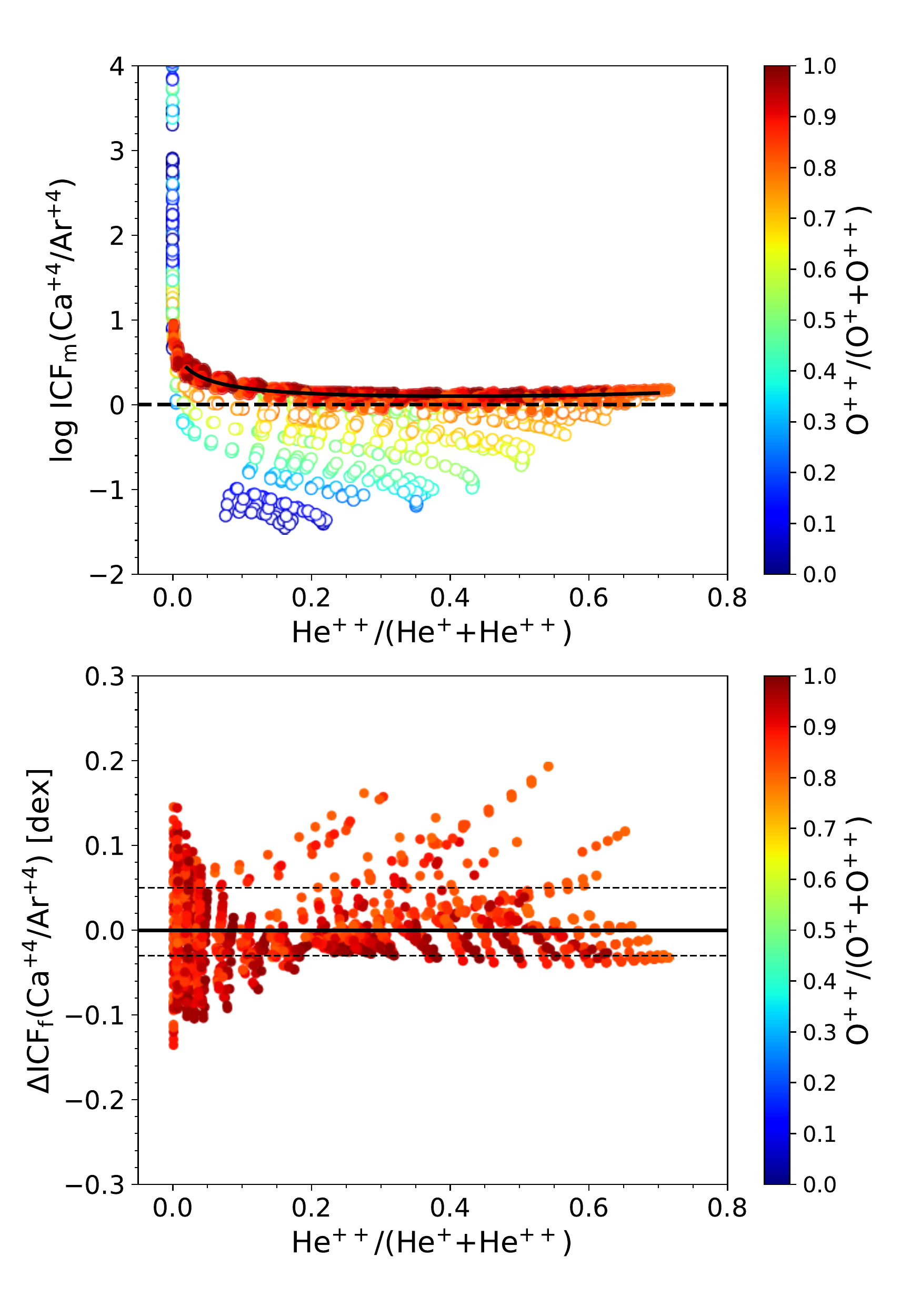}
\caption{{\it Upper panel:} Values of ICF$_{\rm m}$(Ca$^{+4}$/Ar$^{+4}$) as a function of He$^{++}$/(He$^{+}$+He$^{++}$) for our photoionization models. The dashed line represents the ICF proposed by \citet{Bohigas2013}. {\it Lower panel:} Values of $\Delta$ICF$_{\rm f}$(Ca$^{+4}$/Ar$^{+4}$) as a function of He$^{++}$/(He$^{+}$+He$^{++}$) for the photoionization models with \ome~$\geq 0.8$ and \ups~$\geq 0.02$. The solid black line represents where ICF$_{\rm m}$(Ca$^{+4}$/Ar$^{+4}$) = ICF$_{\rm f}$(Ca$^{+4}$/Ar$^{+4}$). The dashed lines represent the uncertainty range adopted for this ICF. The colorbar located on the right side of both panels runs from low to high values of O$^{++}$/(O$^{+}$+O$^{++}$).}
\label{fig:Ca4Ar4}
\end{figure}

As in the case of potassium, we derived a modified ICF for calcium from the one proposed by \citet{Bohigas2013}, Ca/Ar = Ca$^{+4}$/Ar$^{+4}$, to be used when only [\ion{Ca}{v}] lines are observed:
\begin{equation}\label{ICFCa4}
\log {\rm ICF}_{\rm f}({\rm Ca}^{+4}/{\rm Ar}^{+4}) = 0.31\upsilon^{(-0.16+3.40\upsilon)}.
\end{equation}
This expression (represented with a solid line in Figure~\ref{fig:Ca4Ar4}) is valid \ome~$\geq0.8$ and \ups~$\geq0.02$. The lower panel of Figure~\ref{fig:Ca4Ar4} shows that the typical uncertainties associated with this ICF are $^{+0.05}_{-0.03}$ dex.\\

Table~\ref{tab:2} summarizes all the analytical expressions for the ICFs derived here, the ions that have to be observed, the associated uncertainties, and the range of validity of each ICF. We have tested the validity of the ICFs derived here on the additional families of models presented by \citet{DI2014}. These families include models with blackbody and Rauch atmospheres, different metallicities (Z$_{\odot}$/2, Z$_{\odot}$, and 2$\times$Z$_{\odot}$), different nebular density distributions (constant and gaussian), different mass cuts (including matter- and radiation-bounded models), and with and without dust grains. We found that our ICFs are valid for all the models.

\begin{table*}
\centering
\caption{Analytical expressions of all the ICFs proposed here.}
\begin{tabular}{llllll}
    \hline 
    Element & Abundance & Observed & $\log$(ICF) & Range of  & Uncertainties \\
            & ratio     & ions     &             & validity  & [dex] \\
    \hline         
Na   & Na/O & Na$^{++}$/(O$^{+}$+O$^{++}$)  & $0.06 - 0.45\upsilon + 1.63\upsilon^2 - 1.30\upsilon^3$ & \ups~$\geq0.02$, \ome~$\geq0.6$  &  $\pm0.02$  \\
         & Na/O & Na$^{++}$/O$^{+}$  & $0.15/(\omega^{0.80}-1.08)$ & \ups~$\geq0.02$ &  $\pm0.05$  \\
         & Na/O & Na$^{+3}$/O$^{++}$  & $0.11/(0.07 + 1.59\upsilon)$ & \ups~$\geq0.05$ (\ups~$>0.5$)                    &  $^{+0.2}_{-0.4} (\pm0.1)$\\
         & Na/O & (Na$^{++}$+Na$^{+3}$)/(O$^{+}$+O$^{++}$)  & $0.09-0.17 \upsilon^{(0.36-1.19\upsilon)}$  & \ups~$\geq0.02$, \ome~$\geq0.6$ & $^{+0.02}_{-0.03}$   \\
         & Na/O & (Na$^{++}$+Na$^{+3}$)/(O$^{+}$+O$^{++}$)  & $-0.26+1.03\omega-1.77\omega^{2}+0.91\omega^3$  & \ups~$\geq0.02$ & $^{+0.03}_{-0.15}$   \\ \hline
K & K/Ar & K$^{+3}$/Ar$^{+3}$   & $0.343 - 1.12\upsilon + 2.15\upsilon^2 - 1.09\upsilon^3$  & \ups~$\geq0.02$ & $\pm0.05$  \\ \hline
Ca  & Ca/Ar & Ca$^{+4}$/Ar$^{+4}$ &  $0.31\upsilon^{(-0.16+3.40\upsilon)}$ & \ups~$\geq0.02$, \ome~$\geq0.8$ & $^{+0.05}_{-0.03}$ \\
\hline
\end{tabular}
\label{tab:2}
\end{table*}

\section{Using our ICFs}
\label{sec:abunds}

\subsection{Observational sample}
\label{sec:ObsSample}
We use our ICFs to determine the chemical abundances in a sample of PNe from the literature. In Table~\ref{tab:1} we summarize the results of our literature survey for sodium, potassium and calcium lines identified in the optical and infrared spectra of PNe. In columns 2--4 we present the lines identified, the number of PNe where each line has been identified, and the references from which we obtained the corrected intensities, respectively.  

\begin{table}
\centering
\caption{Emission lines of sodium, potassium, and calcium observed in PNe (from the literature).}
\begin{tabular}{llll}
    \hline 
    Element & Lines & No. of & Reference \\
     &  & PNe &  \\
    \hline         
Sodium  & {\fnaiv} $\lambda$3242 \AA  & 4      & 1, 2, 3, 4   \\
        & {\fnaiv} $\lambda$3362 \AA  & 4      & 1, 3, 4, 5 \\
        & {\fnavii} $\lambda$4.68 $\mu$m & 1    & 6 \\
        & {\fnaiii} $\lambda$7.32 $\mu$m&  3    & 7, 8, 9  \\
        & {\fnavi} $\lambda$8.61 $\mu$m & 2     & 6, 8  \\
        & {\fnaiv} $\lambda$9.04 $\mu$m & 2     & 6, 8  \\
        & {\fnavi} $\lambda$14.39 $\mu$m & 1    & 6 \\
        & {\fnaiv} $\lambda$21.29 $\mu$m & 1    & 6 \\ \hline
Potasium& {\fkv} $\lambda$4123 \AA    & 2       & 1, 10 \\
        & {\fkv} $\lambda$4163 \AA    & 6       &  1, 2, 5, 10 \\ 
        & {\fkiv} $\lambda$4511 \AA   & 2       &  1, 2 \\
        & {\fkvi} $\lambda$5602 \AA   & 1       &  1 \\
        & {\fkiv} $\lambda$6102 \AA   & 41      &  1, 2, 3, 4, 5, 10,\\
        &                             &         &  11, 12, 13, 14, \\ 
        &                             &         &  15, 16, 17, 18,  \\
        &                             &         &  19, 20\\
        & {\fkvi} $\lambda$6229 \AA   & 4       &  1, 5, 13, 17, 18 \\ 
        & {\fkiv} $\lambda$6796 \AA   & 20      &  1, 3, 5, 6, 7, 10, \\
        &                             &         &  15, 18, 19 \\ 
        & {\fkvii} $\lambda$3.19 $\mu$m & 1     & 6 \\
        & {\fkiii} $\lambda$4.62 $\mu$m & 1     & 6 \\
        & {\fkvi} $\lambda$5.58 $\mu$m & 1     & 6 \\
        & {\fkiv} $\lambda$5.98 $\mu$m & 1      & 6 \\
        & {\fkvi} $\lambda$8.83 $\mu$m & 1      & 6 \\
        & {\fkiv} $\lambda$15.38 $\mu$m& 1      & 6  \\ \hline
Calcium & {\fcav} $\lambda$5309 \AA   & 7      &  1, 2, 4, 10, \\
        &                              &       & 11, 13, 18\\                             
        & {\fcav} $\lambda$6087 \AA   & 10     &  2, 4, 10, 11, \\
        &                             &        & 12, 13, 18\\    
        & {\fcaii} $\lambda$7292 \AA  & 5      & 10, 15  \\
        & {\fcaii} $\lambda$7324 \AA  & 1      & 15 \\
        & {\fcaiv} $\lambda$3.21 $\mu$m& 1     & 6   \\
        & {\fcav} $\lambda$4.16 $\mu$m& 1      & 6   \\
        & {\fcaviii} $\lambda$6.15 $\mu$m& 1      & 6   \\
        & {\fcav} $\lambda$11.49 $\mu$m& 1   & 6 \\
\hline
    \end{tabular}
\begin{flushleft}
{\bf References}--- (1) \citet{GR2015}; (2) \citet{Hyung1994}; (3) \citet{Hyung2001}; (4) \citet{Hyung1995}; (5) \citet{Tsamis2003}; (6) \citet{BeintemaPott1999}; (7) \citet{Pottasch2009}; (8) \citet{Pottasch2007}; (9) \citet{Pottasch2003b}; (10) \citet{GR2018}; (11) \citet{Bohigas2013}; (12) \citet{Kwitter2001};  (13) \citet{Aller1978}; (14) \citet{Wesson2004}; (15) \citet{GR2012}; (16) \citet{Wesson2018}; (17) \citet{Pottasch2003a};    (18) \citet{Aller1999}; (19) \citet{GR2009};    (20) \citet{Liu2000} 
\end{flushleft}\label{tab:1}
\end{table}

In total we have found emission lines from these three elements in 39 PNe. For four of them we have used observations from two sources: IC~2165 (\citealt{Hyung1994, Bohigas2013}), IC~5217 (\citealt{Hyung2001, Kwitter2001}), NGC~6302 (\citealt{Aller1978, Tsamis2003}), and NGC~6886 (\citealt{Hyung1995, Kwitter2001}). And for NGC~2867 we have used the observations of two different velocity components in the PN observed by \citealt{GR2009}. We decided to used both spectra for each of these PNe because the characteristics of the observations and the ions observed are different. This serves to compare the results. 

In the case of NGC~6153, we used the optical observations from \citet{Liu2000} and the infrared {\fnaiii} $\lambda$7.32 $\mu$m line from \citet{Pottasch2003b}. For Hu~1-2 we used the intensities from the combined red- and blue-shifted line profiles of the North blob reported by \citet{Pottasch2003a}. 

As we mentioned in Section~\ref{sec:ICF}, we have not included the data of NGC~6302 from \citet{BeintemaPott1999} although they reported several Na, K, and Ca lines because most of these lines have only been detected in this object.

\subsection{Physical conditions}\label{physcond}
We started from the intensity ratios with respect to H$\beta$ already corrected from interstellar extinction and we adopted the uncertainties in the line fluxes provided by the authors. When uncertainties are not reported, we adopted a relative uncertainty of 25\% for lines with $I_{\lambda}$/$I_{H\beta}~<0.01$, 10\% for lines with 0.01 < $I_{\lambda}$/$I_{H\beta}~<0.1$, 8\% for lines with 0.1 < $I_{\lambda}$/$I_{H\beta}~<0.2$, and 6\% for lines with $I_{\lambda}$/$I_{H\beta}~>0.2$.

We assumed that the PNe can be characterized by two electron temperatures (\temp) and one electron density (\dens). First, we used the intensity ratios [\ion{O}{ii}] $\lambda$3726/$\lambda$3729, [\ion{S}{ii}] $\lambda$6731/$\lambda$6716, [\ion{Cl}{iii}] $\lambda$5538/$\lambda$5518, and [\ion{Ar}{iv}] $\lambda$4740/$\lambda$4711 to derive an average \dens\ for each PN. Then, we used this averaged \dens\ to compute two \temp's for each PN using the intensity ratios [\ion{N}{ii}] $\lambda$5755/$\lambda$6584 and [\ion{O}{iii}] $\lambda$4363/$\lambda$4959 or [\ion{O}{iii}] $\lambda$4363/$\lambda$5007. If one of the two \temp's is not available we used the other one to compute all the ionic abundances. 

All the calculations have been made with PyNeb \citep{Luridiana2015} and the adopted atomic data are shown in Table~\ref{tab:4}. Some of them were taken from the version 8 {\sc{CHIANTI}} database (\citealt{ChiantiOri, ChiantiVersion}).

\begin{table*}
\caption{Atomic data.}
\begin{tabular}{lll}
\hline
Ion                             & Transition Probabilities        & Collisional Strenghts \\ \hline
N$^{+}$                        & \citet{FFT04}                   & \citet{T11}          \\ \hline
O$^{+}$                         & \citet{FFT04}                   & \citet{Kal09}           \\ \hline
O$^{++}$                         & \citet{FFT04}                   & \citet{SSB14}               \\ \hline
Na$^{++}$                        & \citet{Witthoeft2007}              & \citet{Witthoeft2007}            \\ \hline
Na$^{+3}$                       & \citet{Landi2005}            & \citet{Butler1994}             \\ \hline
Na$^{+5}$                       & \citet{Storey2000}            & \citet{Zhang1996}             \\ \hline
S$^{+}$                        & \citet{VVF96}                   & \citet{TZ10}           \\
\multicolumn{1}{l}{}           & \citet{M82a}                    &                       \\ \hline
Cl$^{++}$                        & \citet{M83}                   & \citet{BZ89}           \\ \hline
Ar$^{+3}$                        & \citet{M82a}                   & \citet{RB97}           \\ \hline
K$^{+3}$                        & \citet{M83}                   & \citet{GMZ95}           \\
\multicolumn{1}{l}{}           & \citet{KS86}                    &                       \\ \hline
K$^{+4}$                        & \citet{M82a}            & \citet{Wilson2001}           \\ \hline
K$^{+5}$                        & \citet{Mendoza1982}            & \citet{GMZ95}           \\ \hline
Ca$^{+4}$                       & \citet{Biemont1986}            & \citet{GMZ95}           \\ \hline
\end{tabular}
\label{tab:4}
\end{table*}

To obtain the final values of \temp\ and \dens\ we propagated the uncertainties in the intensity ratios by running 300 Monte Carlo simulations. For each intensity ratio, we generated a Gaussian distribution centered in the observed value and with a sigma equal to the flux uncertainty. Then, the \temp\ and \dens\ were computed for each Monte Carlo run. To obtain the final values, we computed the medians of each distribution and the associated uncertainties, which are those derived from the 16 and 84 percentiles, that define a confidence interval of 68 per cent. The final values obtained for the sample of PNe are presented in Table~\ref{tab:3}.

The physical conditions that we derived are in general consistent within uncertainties with those reported in the literature. The differences can be explained as due to: the adopted diagnostic ratios used, the \temp/\dens\ assumed, the atomic data adopted, and the method used to derive the final values. Here we use the median value of the temperature/density distribution as the final value. In the case of the \dens, the distribution is obtained from averaging the values computed from each run of the Monte Carlo simulation. A crucial factor involved in the discrepancies between our \temp([\ion{N}{II}]) and the ones from the literature is the correction of the recombination contribution to the [N II] $\lambda$5755 line, not considered in this work.

We could not explain the difference of 1200 K between our \temp([\ion{N}{II}]) and the one reported by \citet{Kwitter2001} for M1-80. If we use the atomic data and electron density reported by these authors we still obtain the same difference of 1200 K.

For the five PNe for which we have two available spectra from different authors we can compare the physical conditions derived. We found reasonable agreement between the results obtained from both observations of IC~2165, IC~5217 and NGC~6886, with differences within the uncertainties.  However, for  NGC~6302 and NGC~2867 we found differences of 2300 and 700 K between both estimates of \temp([\ion{O}{III}]), respectively.

\subsection{Ionic abundances}\label{sec:ionic}
We only computed the ionic abundances involved in the calculations of Na, K, and Ca abundances. We adopted \temp([\ion{N}{ii}]) for the calculation of O$^{+}$/H$^{+}$ and \temp([\ion{O}{III}]) for He$^{+}$/H$^{+}$, He$^{++}$/H$^{+}$, O$^{++}$/H$^{+}$, Na$^{++}$/H$^{+}$, Na$^{+3}$/H$^{+}$, Ar$^{++}$/H$^{+}$, Ar$^{+3}$/H$^{+}$, Ar$^{+4}$/H$^{+}$, K$^{+3}$/H$^{+}$, K$^{+4}$/H$^{+}$, K$^{+5}$/H$^{+}$, and  Ca$^{+4}$/H$^{+}$.

The values of He$^{+}$/H$^{+}$ were estimated from the emission line \ion{He}{i} $\lambda$5876 and using the effective recombination coefficients of \citet{SH95} for \ion{H}{i} and those provided by \citet{Pal12, Pal13} for \ion{He}{i} that include corrections for collisional excitation and self-absorption effects. The He$^{++}$/H$^{+}$ values were computed with the \ion{He}{ii} $\lambda$4686 line (in those PNe where the line is observed) and using the effective recombination coefficients given by \citet{SH95}.

We computed all the ionic abundances for each Monte Carlo run and adopted as the final values the median of each abundance distribution. The associated uncertainties are those computed from the 16 and 84 percentiles, that define a confidence interval of 68 per cent. All the ionic abundances are listed in Table~\ref{tab:5} and Table~\ref{tab:6}. We derived  Na$^{++}$/H$^{+}$ for three PNe, Na$^{+3}$/H$^{+}$ for six PNe, K$^{+3}$/H$^{+}$ for 36 PNe, K$^{+4}$/H$^{+}$ for six PNe, K$^{+5}$/H$^{+}$ for three PNe, and Ca$^{+4}$/H$^{+}$ for 11 PNe.

In general our ionic abundances are in agreement within the uncertainties with those reported in the literature and the differences are up to $\sim0.4$ dex. There are some exceptions where we found highest differences: in O$^{+}$/H$^{+}$ for Fg~1, NGC~6537, and NGC~6790 (0.8, 0.6, and 1.4 dex differences, respectively), in K$^{+3}$/H$^{+}$ for IC~5217 and the K$^{+3}$/H$^{+}$ reported by \citet{Hyung1994} for IC~2165 (0.4 and 0.5 dex, respectively), in K$^{+4}$/H$^{+}$ for H~1-50 (0.4 dex), in K$^{+5}$/H$^{+}$ for Hu~1-2 (0.6 dex), and in the Ca$^{+4}$/H$^{+}$ derived by \citet{Hyung1994} for IC~2165 (0.5 dex). The reasons behind the discrepancies are the different atomic data, differences in the physical conditions adopted and the different lines used in the calculations. For example, we only use the [\ion{O}{ii}] $\lambda\lambda$3727,29 lines to derive the O$^{+}$/H$^{+}$ abundances whereas some authors also used the [\ion{O}{ii}] $\lambda\lambda$7320,30 lines. There are one case that we cannot explain: the Ar$^{++}$/H$^{+}$ derived by \citet{Tsamis2003} for NGC~6818 that is $\sim$1.1 dex lower than ours. We were not able to reproduce this value. 

From the comparison between the ionic abundances derived for the objects with two available spectra we found differences of up to 0.2 dex for NGC~2867 and  NGC~6302, up to 0.3 dex for NGC~6886 and IC~2165, and up to 0.4 dex for IC~5217. 

\subsection{Total abundances}\label{sec:abund}

The total abundances of O and Ar were computed using the ICFs derived by \citet{DI2014} and the total abundances of Na, K, and Ca were computed with the ICFs derived here. We take into account two type of uncertainties: 1) those that arise from the uncertainties in the line intensities and 2) those that are associated with the adopted ICF. The first ones are the only ones considered in most of the abundance calculations but the uncertainties associated with the correction scheme may be significant for Ar, Na, K, and Ca. 

The uncertainties associated with the ICFs were estimated by constructing a uniform distribution for each ICF via 300 Monte Carlo simulations. The central value of such distribution is the ICF computed from the observed degree of ionization, and the upper and lower limits are those related to the uncertainties associated with each ICF (see Table~\ref{tab:2}). Then we compute the total abundances associated to the generated random ICF values. The final value of each abundance is the median of the distribution and the associated uncertainties are given by the 16 and 84 percentiles.

Table~\ref{tab:7} shows the final abundances for all the PNe. We do not provide the total abundance of Na, K, or Ca in those PNe with an estimation of the ionic abundances of these elements but with a value of \ome\  or \ups\  outside the range of validity of the ICFs derived here. The total number of PNe with Na, K, and Ca abundances is 8, 28, and 11, respectively.

If we use the ICF suggested by \citet{Bohigas2013} for K and Ca we would obtain systematically lower abundances by up to 0.6 and 0.4 dex, respectively.  

Our derived total abundances of O, K, and Ar are in general consistent within the uncertainties with the values reported in literature. However, there are some exceptions, particularly for Ar and K abundances. For K, we obtained the largest difference with the one derived by \citet{Pottasch2009} for NGC~6210, which is 0.8 dex lesser than the derived here. As for Na and Ca abundances we find less agreement between our values and the ones from the literature, reaching 0.6 dex (bigger than ours) and 1.1 dex (lesser than ours), respectively for IC~5217 from \citet{Hyung2001} and for NGC~6302 from \citet{Aller1978}.

\begin{figure}
\includegraphics[width=\columnwidth, trim={0 10 0 0}, clip]{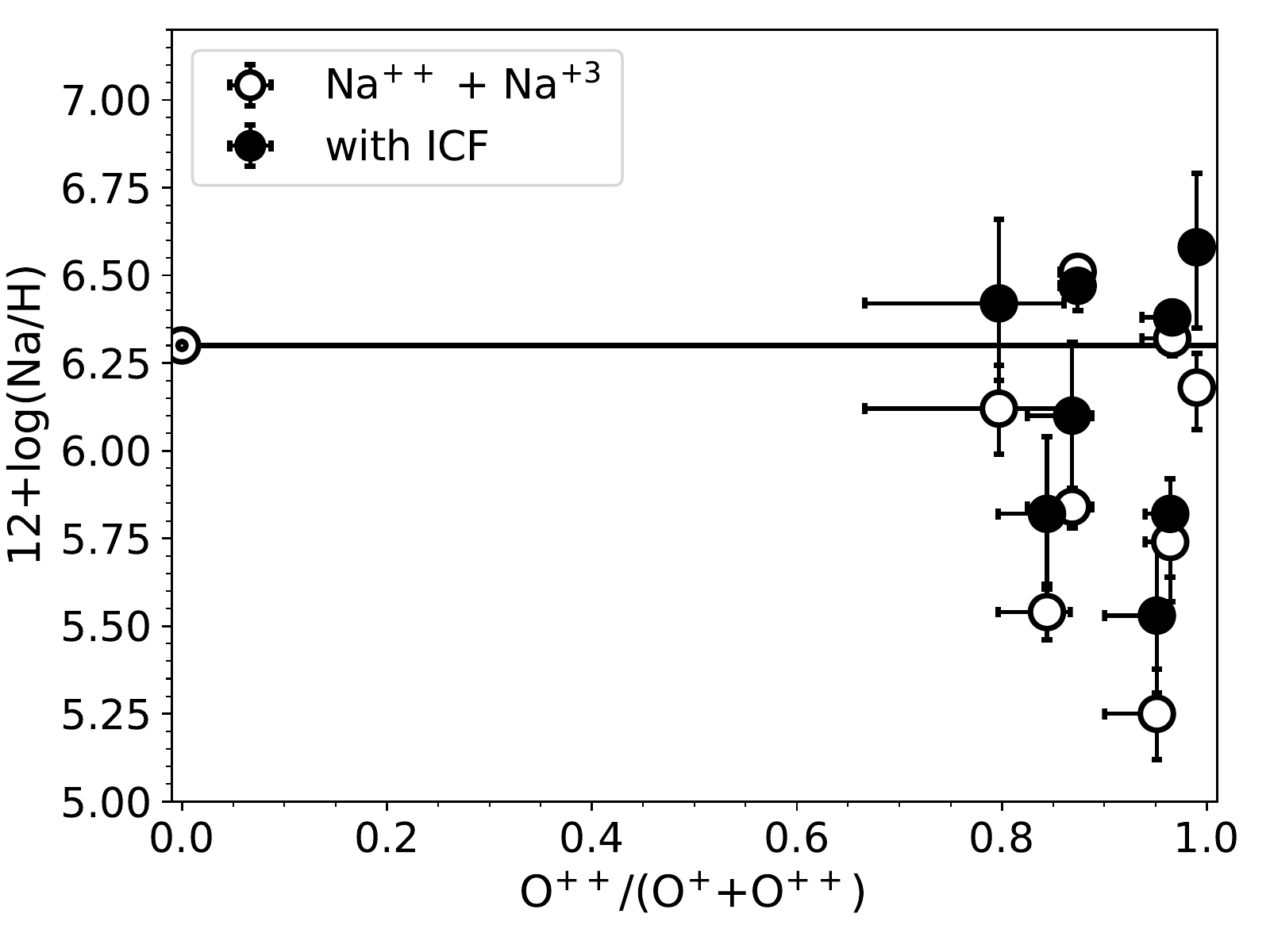}
\caption{Values of Na/H (filled circles) and Na$^{++}$/H$^+$ + Na$^{+3}$/H$^+$ (empty circles) as a function of the degree of ionization given by O$^{++}$/(O$^{+}$+O$^{++}$) for all the PNe. The solid line with the solar symbol at the left represents the solar value by \citet{Lodders2003}.} 
\label{fig:na_obs}
\end{figure}

\begin{figure}
\includegraphics[width=\columnwidth, trim={0 10 0 0}, clip]{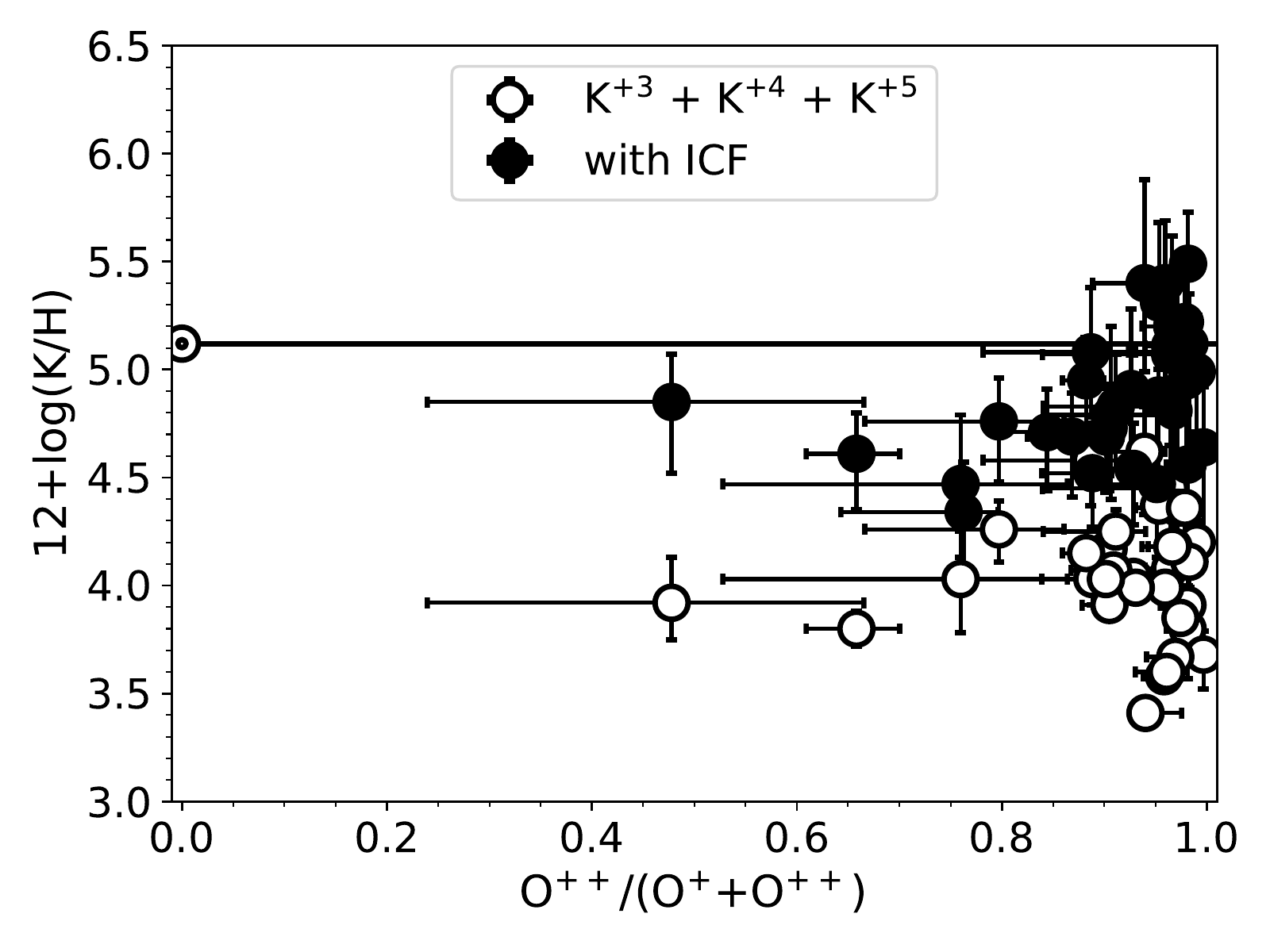}
\caption{Values of K/H (filled circles) and K$^{+3}$ + K$^{+4}$ + K$^{+5}$ (empty circles) as a function of the degree of ionization given by O$^{++}$/(O$^{+}$+O$^{++}$) for all the PNe. The solid line with the solar symbol at the left represents the solar value by \citet{Lodders2003}.} 
\label{fig:k_obs}
\end{figure}

\begin{figure}
\includegraphics[width=\columnwidth, trim={0 10 0 0}, clip]{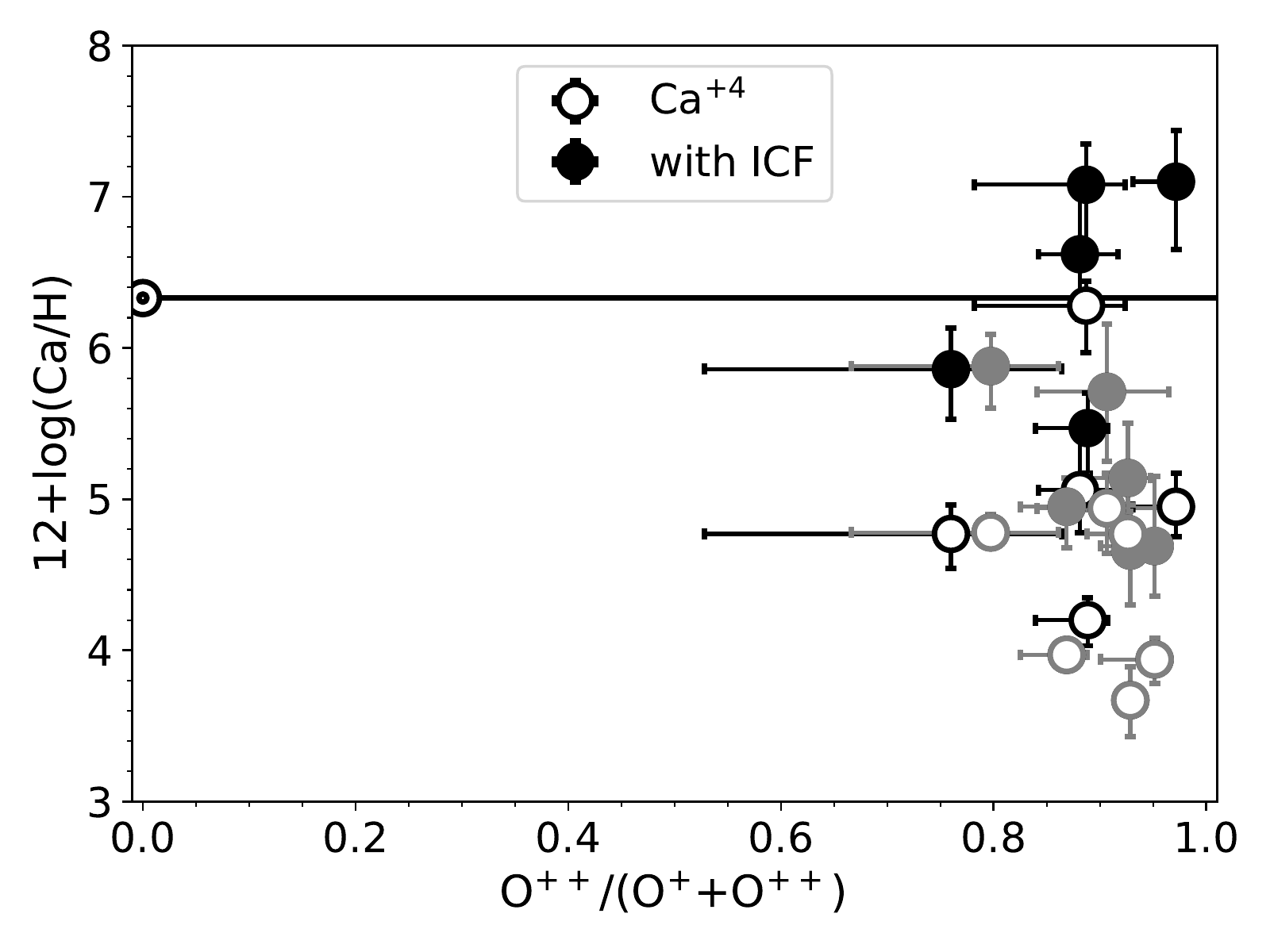}
\caption{Values of Ca/H (filled circles) and Ca$^{+4}$ (empty circles) as a function of the degree of ionization given by O$^{++}$/(O$^{+}$+O$^{++}$) for all the PNe. Grey circles correspond to objects with a blended [Ca~{\sc v}] $\lambda$6086.40 line. The solid line with the solar symbol at the left represents the solar value by \citet{Lodders2003}.} 
\label{fig:ca_obs}
\end{figure}

Figures~\ref{fig:na_obs}-\ref{fig:ca_obs} show the values of the total abundances of Na, K, and Ca computed with our ICFs as a function of the degree of ionization for all the PNe with emission lines of these elements. These figures also show the sum of the ionic abundances of each element observed in each PN. The comparison between both values illustrates that the ICF is significant in many cases and not considering it may lead to incorrect abundances. We can also see in these figures that all the PNe have a relatively high degree of ionization, O$^{++}$/(O$^{+}$+O$^{++}$)~$\gtrsim0.5$ and thus, we cannot test the proposed ICFs in the whole range of degree of ionization. However, since the ions needed to compute Na, K, and Ca abundances have relatively high I.P. we do not expect them to contribute significantly in low ionization PNe and thus, our correction schemes may not be applicable to those objects.

There is no obvious trend between the total abundances of Na, K and Ca and the ionization degree, at least in the range covered by the sample. We computed the Pearson correlation coefficients of the Na/O, K/O and Ca/O abundances as a function of the ionization degree, and obtained coefficients of -0.084, 0.524 and 0.022 with a $p$-value of 0.838, 0.002 and 0.948, respectively. These results indicate that there is a weak but statistically significant relation between the K/O values and the ionization degrees. However, a larger sample of PNe is necessary to find out if our ICF for K is introducing this trend or not.

Figures~\ref{fig:na_obs}-\ref{fig:ca_obs} show that most of the PNe have Na, K, and Ca abundances below the solar value, as expected if these atoms are deposited in dust grains. However, a few PNe have abundances above the solar value. Below, we will discuss this result further.

\section{Depletion factors and discussion}\label{sec:depl}

As we mentioned in Section~\ref{sec:intro} the depletion factor of an element refers to its observed underabundance with respect to the solar abundance, which is generally used as a reference for the total abundance (gas plus dust) of each element. Ele\-ment depletion is presumed to be the consequence of atom condensation into dust grains and it can be computed as: 
\begin{equation}
[{\rm X}/{\rm H}] = \log({\rm X}/{\rm H}) - \log({\rm X}/{\rm H})_{\odot}. 
\end{equation}

\begin{figure}
\includegraphics[width=\columnwidth, trim={5 0 10 0}, clip]{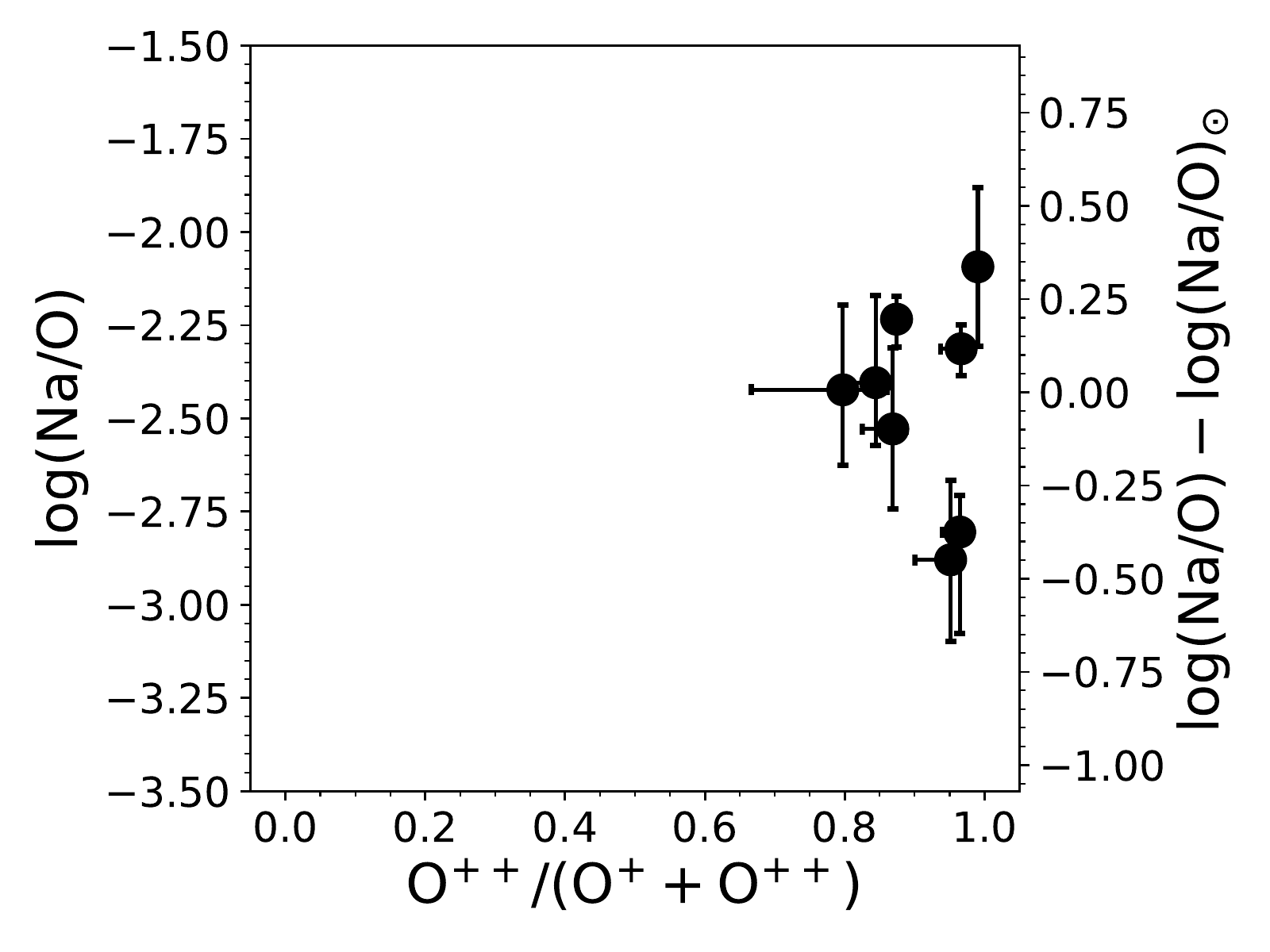}
\caption{Values of Na/O (left axis) and the depletion factor for Na/O (right axis) as a function of the degree of ionization given by O$^{++}$/(O$^{+}$+O$^{++}$) for all the PNe.}
\label{fig:na_obs2}
\end{figure}

\begin{figure}
\includegraphics[width=\columnwidth, trim={5 0 10 0}, clip]{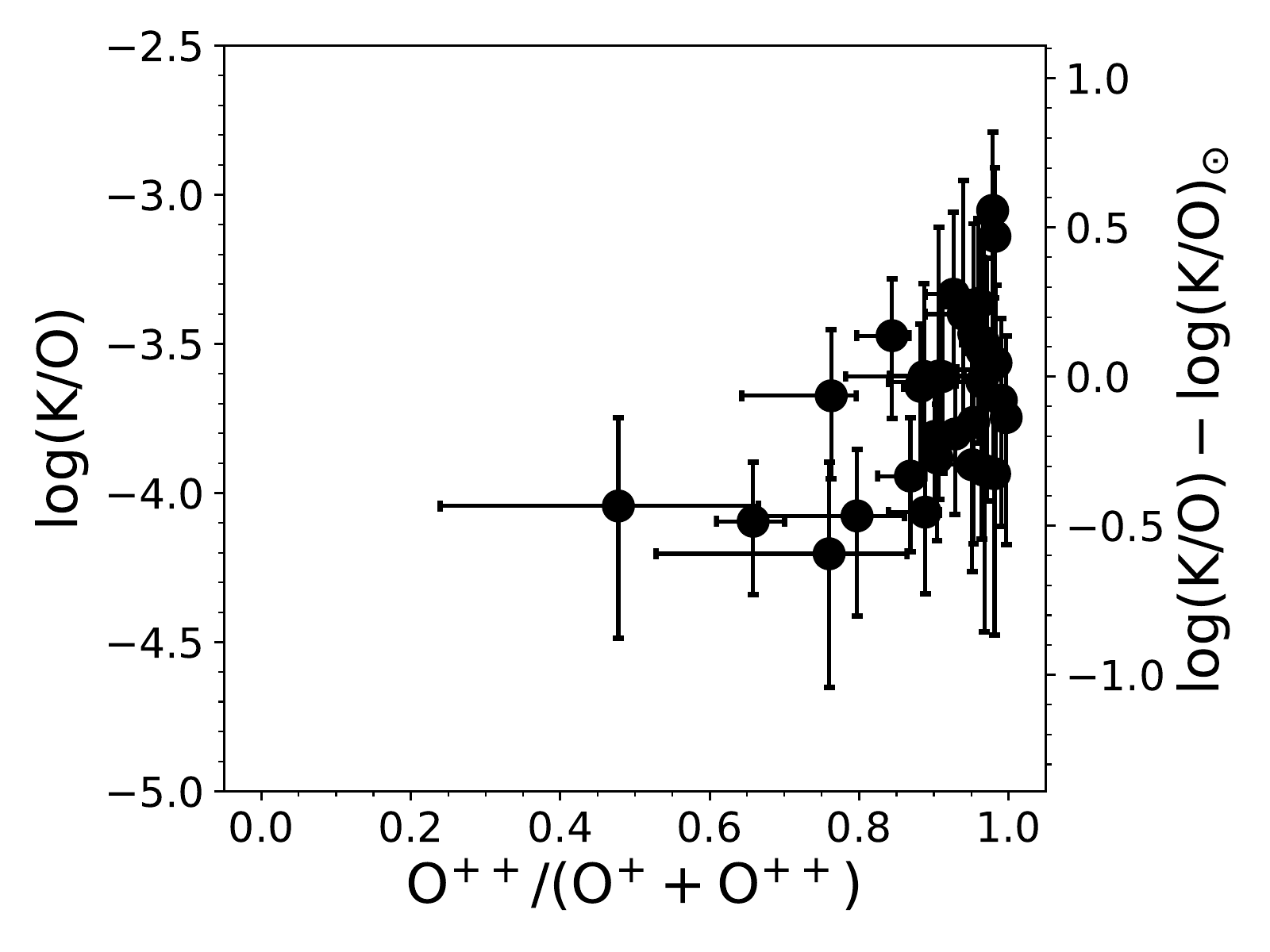}
\caption{Values of K/O (left axis) and the depletion factor for K/O (right axis) as a function of the degree of ionization given by O$^{++}$/(O$^{+}$+O$^{++}$) for all the PNe.}
\label{fig:k_obs2}
\end{figure}

\begin{figure}
\includegraphics[width=\columnwidth, trim={5 0 10 0}, clip]{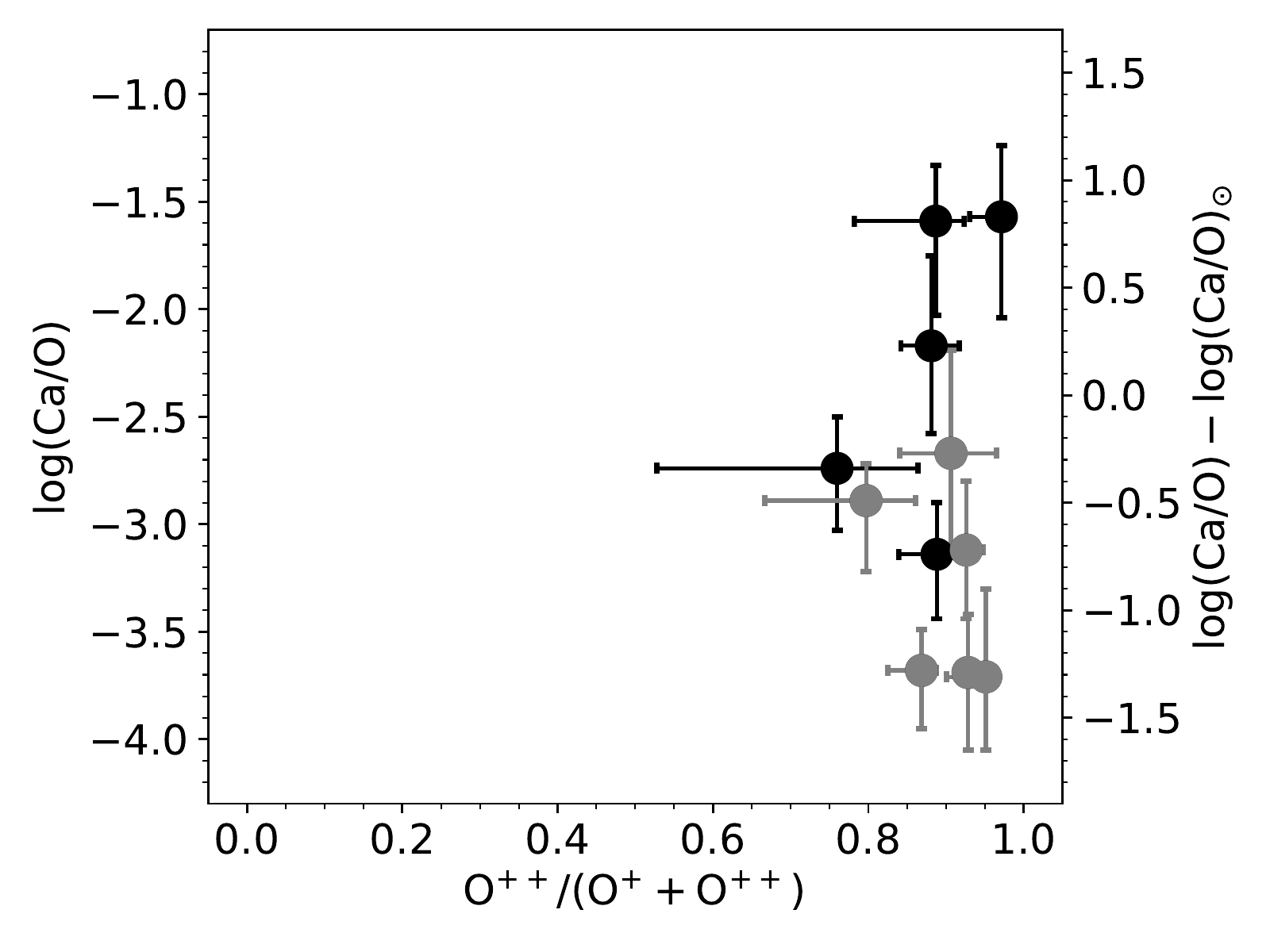}
\caption{Values of Ca/O (left axis) and the depletion factor for Ca/O (right axis) as a function of the degree of ionization given by O$^{++}$/(O$^{+}$+O$^{++}$) for all the PNe. Grey circles correspond to objects with a blended [Ca~{\sc v}] $\lambda$6086.40 line.}
\label{fig:ca_obs2}
\end{figure}

Figures~\ref{fig:na_obs2}-\ref{fig:ca_obs2} show the Na/O, K/O, and Ca/O abundance ratios (left axes) and the depletion factors for Na/O, K/O, and Ca/O (right axes) as a function of the degree of ionization for all the PNe.
This allows us to study if the total abundances of Na, K and Ca (and the depletion factors) show a trend with the degree of ionization of the PNe which, in principle, would indicate that our ICFs are not adequate.

We have used the abundance ratios with respect to oxygen instead of the abundance ratios with respect to hydrogen because the intrinsic value of the first ones are expected to show less variation from one object to another. We used as reference values for the total abundance of Na, K, and Ca the values provided by \citet{Lodders2003}: $\log$(Na/O)$_{\odot}$= $-2.43\pm0.08$, $\log$(K/O)$_{\odot}$= $-3.61\pm0.08$, and $\log$(Ca/O)$_{\odot}$= $-2.40\pm0.08$ (shown in black solid lines in these Figures).

Our values of $\log$(Na/O), $\log$(K/O), and $\log$(Ca/O) range from $-2.88_{-0.22}^{+0.21}$ to $-2.09\pm0.21$, from $-4.20_{-0.45}^{+0.31}$ to $-3.05_{-0.47}^{+0.26}$, and from $-3.71_{-0.34}^{+0.41}$ to $-1.57_{-0.47}^{+0.33}$, respectively. And the values of [Na/O], [K/O], and [Ca/O] range from $-0.45\pm0.20$ to $0.34\pm0.21$, from $-0.60_{-0.45}^{+0.31}$ to $0.56_{-0.47}^{+0.26}$, and from $-1.31_{-0.34}^{+0.41}$ to $0.83_{-0.47}^{+0.33}$ respectively.

As we mentioned above, some PNe have higher abundances than the Sun and thus, the corresponding depletion factors are above zero; which is, in principle, unexpected. The PNe with [Na/O]~$>0$ taking into account the uncertainties in the derived Na/O are Hb~5, IC~5217 from \citet{Hyung2001} and NGC~6153 ([Na/O] = 0.20$^{+0.06}_{-0.08}$, 0.34$\pm0.21$ and 0.12$^{+0.06}_{-0.07}$, respectively). The PNe with [Ca/O]~$>0$ taking into account the uncertainties in the derived Ca/O are M~1-57 and NGC~6884 ([Ca/O] = $0.80^{+0.26}_{-0.44}$ and $0.82^{+0.33}_{-0.47}$, respectively). And finally, the only PN with [K/O]~$>0$ taking into account the uncertainties in the derived K/O is Fg~1 ([K/O] = $0.55^{+0.26}_{-0.47}$).

This result could be partially due to the deposition of oxygen atoms into dust grains. According to \citet{Whittet2010} the values of [Na/O], [K/O], and [Ca/O] should be lowered in $\sim0.15$ dex if oxygen is trapped in oxides and silicates. However, oxygen depletion cannot explain the highest values of [Na/O], [K/O], and [Ca/O] obtained in some PNe (above$\sim0.3$ dex). We also explored the depletion factors using a non-refractory element such as argon, instead of oxygen, and found that the number of objects with positive depletion factors changed for the three elements. When using Ar as the reference element instead of oxygen, we found that there are no objects with positive sodium and potassium depletion factors, taking into account the bigger uncertainties in each case (which were those associated to Ar/H in the case of sodium and those associated to K/H in the case of potassium), while only three objects (M1-57, M1-80 and NGC~6884) have positive calcium depletion factors (with [Ca/Ar] of $0.81\pm0.20$, $0.56_{-0.27}^{+0.17}$ and $0.61_{-0.42}^{+0.29}$, respectively) taking into account the uncertainties of the derived Ar/H ratio, which are bigger to those associated to Ca/H. The differences found in the number of objects with positive depletion factors may be because the oxygen depletion in dust computed by \citet{Whittet2010} is adequate for some specific physical conditions and chemical compositions which that are different to those of these objects, and it may be underestimated in some cases. However, argon abundances are less reliable than oxygen abundances and thus, we preferred to present here the abundances with respect to oxygen. Another explanation for the high abundances derived is that some lines may be misidentified or blended. In fact, the derived Ca abundances for five out of the eleven PNe where Ca$^{+4}$ emission lines are observed is likely overestimated due to a blend reported in [Ca~{\sc v}] $\lambda$6086.40 with the [Fe~{\sc vii}] $\lambda$6086.29 line. These PNe are IC\,2165 from \citet{Hyung1994}, IC\,2165 from \citet{Bohigas2013}, NGC\,6302 from \citet{Aller1978}, NGC\,6537 from \citet{Aller1999},  NGC\,6886 from \citet{Hyung1995}, and NGC\,3918 from \citet{GR2015}. In any case, this blend can only affect highly excited PNe. These five PNe have been included in Figures 11 and 14 (shown in grey circles) because [Ca V] 6086.40 line was the only one available to compute the calcium abundance.

\citet{Field1974} found that the observed depletions correlate with the condensation temperature, $T_{\rm C}$. This parameter is defined as the temperature at which 50 per cent of the atoms are condensed into the solid phase. For the more refractory elements (e.g., Ca, Fe, Si) the trend between depletion and $T_{\rm C}$ is strongest as $T_{\rm C}$ is higher whereas the more volatile elements (e.g., N, S) have very low depletions independently on $T_{\rm C}$ \citep{Whittet2003}. For a solar-system composition gas, the $T_{\rm C}$ of sodium, potassium, and calcium are 958, 1006, and 1517 K, respectively \citep{Lodders2003}. Therefore, we expect calcium to have the highest depletions whereas sodium and potassium (moderately volatile elements) should have low, or not depletion at all.

Figures~\ref{fig:na_obs2}-\ref{fig:ca_obs2} show that this is exactly the case. There are five PNe with [Ca/O]~$\leq-0.5$, only one PN with [K/O]~$\leq-0.5$, and none with [Na/O]~$\leq-0.5$. 
The highest Ca depletions are found for NGC~3918 and both observations of IC~2165, where we obtained that $\sim$ 95\% of the calcium atoms in these nebulae are deposited in the solid grains. NGC~3918 has an almost solar Na abundance and $\sim$ 50\% of its K atoms in the dust phase. As for IC~2165, $\sim$ 40\% of its K atoms are likely condensed in the dust grains. 
The maximum Na depletions are obtained for IC~2165 from \citet{Hyung1994} and NGC~6210, where $\sim$ 60\% of its Na atoms are deposited in the dust phase, while the maximum K depletions were found for both observations of NGC~6886 and NGC~5189, where $\sim$ 70\% of their K atoms is in the dust phase.

Sodium may be synthesized during the H-burning and then brought to the stellar surface through the first dredge-up \citep{Mowlavi1999}. Since the Na is produced from initial Ne, its origin is secondary. \citet{Mowlavi1999} proposed a primary production of Na in AGB stars from the Ne produced in the He-burning shell and its subsequent transformation into Na via the H-burning shell. The material is then mixed and transported into the stellar surface via the third dredge-up. The production of Na competes with its depletion into dust grains. We can test these two scenarios by comparing the Na abundances and the C/O abundance ratios of our PN sample. If the sodium origin is primary, a positive correlation with C/O is expected. We compiled the values of C/O from \citet{delgadoinglada2014}, except for those of NGC~6886, NGC~6302 from \citet{Tsamis2003}, and Hb~5, which were taken from the references of Table~\ref{tab:1}. We did not find any obvious correlation but a larger sample is needed to make any conclusion (the Pearson correlation coefficient is 0.15 and the p-value is 0.81).

The maximum depletions reached here for calcium are lower than the ones obtained for iron or nickel in PNe: [Fe/O] and [Ni/O], up to $\sim-3$ \citep{delgadoinglada2016}. This is surprising because calcium has a somewhat higher $T_{\rm C}$ (1517 K) than iron (1334 K) and nickel (1353 K). The higher iron and nickel depletions found in some PNe could be related to a higher efficiency of these elements to form grains or to differences in the formation and/or destruction of grains in the PNe involved. However, more spectra of high quality and of more PNe are needed to clarify this issue. 

\section{Conclusions}
\label{sec:summary}
We have derived ionization correction factors (ICFs) for sodium, potassium, and calcium using a large grid of photoionization models derived by \citet{DI2014}. The grid is part of the Mexican Million Models database and it is available for the community. 

We also provide the typical uncertainties associated with $\log$(Na/O), $\log$(K/Ar) and $\log$(Ca/Ar) arising from our ICFs. For each ICF, we define the range of validity, i.e., the range of O$^{++}$/(O$^{+}$+O$^{++}$) and/or He$^{++}$/(He$^{+}$+He$^{++}$) where the total abundance can be safely derived. Outside this region we expect that the error bars are large and thus, the ICFs are very uncertain and we recommend not to use them. In these cases computing a detailed photoionization model is the only possibility to obtain a reliable abundances.

The only previous ICFs in the literature are those derived by \citet{Bohigas2013} for potassium based on K$^{+3}$ abundance and for calcium based on Ca$^{+4}$ abundance. Both ICFs have been significantly improved here since our ICFs are based on realistic photoionization models. This is especially the case for K abundances where the previous ICF systematically underestimates them.   

We tested our ICFs with a sample of 39 PNe with emission lines of some ion of these elements. We were able to compute Na abundances in eight PNe, K abundances in 28 PNe, and Ca abundances in nine PNe. We found no obvious trend between the derived abundances and the degree of ionization but more PNe are required to obtain statistically significant results and to confirm that our ICFs are not introducing an artificial bias in the results.

Two PNe have a depletion factor above zero within the uncertainties, which cannot be explained due to oxygen atoms deposition onto dust grains. Emission line miss-identifications or blends may explain this result. The depletions found for these three elements with oxygen as the reference element, range from $-0.45\pm0.20$ to $0.34\pm0.21$ for [Na/O], from $-0.60_{-0.45}^{+0.31}$ to $0.56_{-0.47}^{+0.26}$ for [K/O], and from $-1.31_{-0.34}^{+0.41}$ to $0.83_{-0.47}^{+0.33}$ for [Ca/O]. These numbers imply that some of the studied PNe have up to 65\%, 75\%, or 95\% of their Na, K, or Ca atoms condensed into dust grains, respectively. Our results are consistent with expectations since calcium (with the highest condensation temperature) shows the highest depletions.  

\section*{Acknowledgements}
 
We thank an anonymous referee for his/her detailed comments that helped to improve the paper. We also thank C. Morisset and G. Stasi\'nska who read previous drafts of this paper and provided very valuable comments. AA and GD-I acknowledge support from PAPIIT (DGAPA-UNAM) grant no. IA$-$101517. AA thanks CONACyT for her Master and PhD scholarship (825508). GD-I received partial support from CONACyT grant 241732. JG-R acknowledges support from an Advanced Fellowship from the Severo Ochoa excellence program (SEV-2015-0548) and support from the State Research Agency (AEI) of the Spanish Ministry of Science, Innovation and Universities (MCIU) and the European Regional Development Fund (FEDER) under grant AYA2017-83383-P. JG-R also acknowledges support under grant P/308614 financed by funds transferred from the Spanish Ministry of Science, Innovation and Universities, charged to the General State Budgets and with funds transferred from the General Budgets of the Autonomous Community of the Canary Islands by the MCIU.

\bibliographystyle{mnras}

\begin{table*}
\centering
\caption{Physical conditions.}
\begin{threeparttable}
\begin{tabular}{lllll}
\hline         
Name    & {\temp}([\ion{N}{ii}]) & {\temp}([\ion{O}{iii}]) & {\dens} & Ref. \\
\hline
Fg\,1	&	---	&	10900$\pm300$	&	700$\pm500$	&	16\\
H\,1-40	&	13300$_{-1400}^{+1000}$	&	10100$_{-400}^{+300}$	&	9200$_{-3300}^{+9200}$	&	10\\
H\,1-50	&	12000$_{-600}^{+500}$	&	11100$\pm300$	&	8800$_{-1700}^{+2200}$	&	10\\
Hb\,4	&	9700$\pm700$	&	9900$\pm300$	&	7500$_{-2400}^{+6800}$	&	15\\
Hb\,5	&	---	&	12900$_{-400}^{+500}$	&	8100$_{-3200}^{+8500}$	&	8\\
He\,2-73	&	11200$\pm500$	&	11800$\pm300$	&	10200$_{-1900}^{+3900}$	&	10\\
Hen\,2-86	&	10300$\pm700$	&	8400$\pm200$	&	22000$_{-3700}^{+5300}$	&	15\\
Hu\,1-2	&	13200$_{-1000}^{+900}$	&	20400$\pm1200$	&	7800$_{-1900}^{+3000}$	&	17\\
IC\,2165	&	12100$\pm300$	&	13800$\pm200$	&	4200$\pm400$	&	11\\
IC\,2165	&	12700$_{-1800}^{+1500}$	&	14300$\pm600$	&	4900$_{-1500}^{+2200}$	&	2\\
IC\,4191	&	11800$_{-400}^{+500}$	&	9900$\pm200$	&	10000$\pm1200$	&	5\\
IC\,5217	&	12800$_{-1800}^{+1900}$	&	10800$\pm400$	&	5100$_{-1500}^{+2900}$	&	3\\
IC\,5217	&	12900$_{-1400}^{+1300}$	&	11200$_{-500}^{+600}$	&	8300$_{-3600}^{+8500}$	&	12\\
NGC\,2022	&	---	&	14800$_{-400}^{+500}$	&	1300$\pm400$	&	5\\
NGC\,2867-1	&	11500$\pm400$	&	12200$\pm300$	&	4100$_{-600}^{+700}$	&	19\\
NGC\,2867-2	&	11300$\pm400$	&	11500$\pm300$	&	3500$_{-400}^{+500}$	&	19\\
NGC\,3242	&	12200$\pm1300$	&	11700$\pm300$	&	1900$_{-300}^{+400}$	&	5\\
NGC\,3918	&	10900$\pm500$	&	12700$\pm400$	&	6200$_{-800}^{+1800}$	&	1\\
NGC\,5189	&	9500$_{-300}^{+400}$	&	11500$\pm300$	&	1300$_{-300}^{+200}$	&	15\\
NGC\,5882	&	10400$\pm300$	&	9300$\pm200$	&	5000$_{-600}^{+800}$	&	5\\
NGC\,6153	&	10200$_{-1100}^{+1000}$	&	9000$\pm300$	&	3900$_{-1100}^{+2200}$	&	9,20\\
NGC\,6210	&	10800$\pm700$	&	9600$\pm400$	&	5300$_{-1400}^{+2200}$	&	7\\
NGC\,6302	&	13300$\pm700$	&	18300$_{-700}^{+800}$	&	14300$_{-1900}^{+2300}$	&	5\\
NGC\,6302	&	15800$_{-3100}^{+1700}$	&	16000$_{-800}^{+700}$	&	12800$_{-5200}^{+19800}$	&	13\\
NGC\,6309	&	10000$_{-2300}^{+2500}$	&	11900$\pm600$	&	3300$_{-2400}^{+7700}$	&	12\\
NGC\,6369	&	13200$\pm700$	&	10700$_{-300}^{+200}$	&	4400$_{-600}^{+1000}$	&	15\\
NGC\,6537	&	17000$_{-1600}^{+2000}$	&	16400$_{-1800}^{+2200}$	&	18000$_{-4500}^{+7600}$	&	18\\
NGC\,6543	&	9900$_{-800}^{+700}$	&	7800$\pm100$	&	6200$_{-1600}^{+3000}$	&	14\\
NGC\,6572	&	14300$_{-1800}^{+1900}$	&	10100$\pm500$	&	10500$_{-4700}^{+12200}$	&	12\\
NGC\,6790	&	19800$_{-3200}^{+3400}$	&	12600$_{-600}^{+800}$	&	6300$_{-4800}^{+8900}$	&	12\\
NGC\,6818	&	11000$_{-300}^{+400}$	&	13400$_{-300}^{+400}$	&	2000$\pm300$	&	5\\
NGC\,6884	&	11300$_{-1600}^{+1700}$	&	10600$\pm500$	&	7800$_{-3900}^{+7500}$	&	12\\
NGC\,6886	&	10200$_{-900}^{+700}$	&	11600$_{-500}^{+400}$	&	9900$_{-2900}^{+8000}$	&	4\\
NGC\,6886	&	11300$_{-1300}^{+1100}$	&	12200$_{-700}^{+800}$	&	9500$_{-4600}^{+12100}$	&	12\\
NGC\,7026	&	9900$\pm600$	&	8700$_{-700}^{+500}$	&	4500$_{-1900}^{+4200}$	&	12\\
M\,1-33	&	9300$_{-300}^{+400}$	&	8900$\pm200$	&	6200$_{-1200}^{+1500}$	&	10\\
M\,1-50	&	10200$_{-1600}^{+1000}$	&	10500$\pm500$	&	10900$_{-6200}^{+26000}$	&	12\\
M\,1-54	&	8700$\pm700$	&	10100$_{-500}^{+400}$	&	9000$_{-4800}^{+10700}$	&	12\\
M\,1-57	&	11700$_{-900}^{+1100}$	&	12200$\pm600$	&	6000$_{-2600}^{+4300}$	&	12\\
M\,1-60	&	9700$_{-400}^{+300}$	&	8800$\pm200$	&	12800$_{-2300}^{+3400}$	&	10\\
M\,1-61	&	12600$_{-1000}^{+900}$	&	9100$\pm200$	&	23500$_{-4700}^{+5800}$	&	15\\
M\,1-80	&	11200$\pm700$	&	9800$\pm400$	&	1000$_{-500}^{+700}$	&	12\\
M\,2-31	&	10900$\pm400$	&	9900$_{-200}^{+300}$	&	7900$_{-1300}^{+1800}$	&	10\\
PC\,14	&	10200$\pm400$	&	9300$\pm200$	&	4200$_{-600}^{+800}$	&	15\\
\hline         
\end{tabular}
\begin{tablenotes}
\item[] {\bf Note:} References are the same as those in Table~\ref{tab:1}.\\
\end{tablenotes}
\end{threeparttable}
\label{tab:3}
\end{table*}

\begin{table*}
\centering
\caption{Ionic abundances of sodium, potassium and calcium ions in 12$+\log$(X$^{+i}$/H$^{+}$).}
\begin{threeparttable}
\begin{tabular}{llllllll}
\hline         
Name        & Na$^{++}$/H$^{+}$     & Na$^{+3}$/H$^{+}$     & K$^{+3}$/H$^{+}$  & K$^{+4}$/H$^{+}$   & K$^{+5}$/H$^{+}$ & Ca$^{+4}$/H$^{+}$\\
\hline
Fg\,1	&	---	&	---	&	4.36$_{-0.18}^{+0.11}$	&	---	&	---	&	---\\
H\,1-40	&	---	&	---	&	3.60$\pm0.16$	&	---	&	---	&	---\\
H\,1-50	&	---	&	---	&	4.26$_{-0.05}^{+0.06}$	&	3.60$_{-0.16}^{+0.11}$	&	---	&	---\\
Hb\,4	&	---	&	---	&	4.25$\pm0.05$	&	---	&	---	&	---\\
Hb\,5	&	6.41$\pm0.04$	&	5.79$\pm0.04$	&	---	&	---	&	---	&	---\\
He\,2-73	&	---	&	---	&	4.03$_{-0.07}^{+0.06}$	&	---	&	---	&	4.20$_{-0.16}^{+0.12}$\\
Hen\,2-86	&	---	&	---	&	3.58$_{-0.06}^{+0.05}$	&	---	&	---	&	---\\
IC\,2165	&	---	&	---	&	4.04$_{-0.05}^{+0.04}$	&	---	&	---	&	3.67$_{-0.25}^{+0.16}$\\
IC\,2165	&	---	&	5.25$_{-0.14}^{+0.13}$	&	4.07$_{-0.14}^{+0.10}$	&	3.95$\pm0.12$	&	---	&	3.94$_{-0.14}^{+0.10}$\\
IC\,4191	&	---	&	---	&	4.37$\pm0.06$	&	---	&	---	&	---\\
IC\,5217	&	---	&	6.18$_{-0.10}^{+0.11}$	&	4.20$_{-0.09}^{+0.08}$	&	---	&	---	&	---\\
IC\,5217	&	---	&	---	&	3.80$_{-0.23}^{+0.17}$	&	---	&	---	&	---\\
NGC\,2022	&	---	&	---	&	4.43$_{-0.08}^{+0.07}$	&	3.95$\pm0.06$	&	---	&	---\\
NGC\,2867-1	&	---	&	---	&	4.03$_{-0.07}^{+0.06}$	&	---	&	---	&	---\\
NGC\,2867-2	&	---	&	---	&	4.15$_{-0.09}^{+0.07}$	&	---	&	---	&	---\\
NGC\,3242	&	---	&	---	&	4.20$_{-0.05}^{+0.06}$	&	3.41$\pm0.04$	&	---	&	---\\
NGC\,3918	&	---	&	5.84$\pm0.06$	&	4.31$\pm0.04$	&	3.93$_{-0.05}^{+0.04}$	&	3.01$_{-0.09}^{+0.08}$	&	3.97$_{-0.04}^{+0.05}$\\
NGC\,5189	&	---	&	---	&	3.80$_{-0.07}^{+0.09}$	&	---	&	---	&	---\\
NGC\,5882	&	---	&	---	&	4.11$_{-0.11}^{+0.08}$	&	---	&	---	&	---\\
NGC\,6153	&	6.32$\pm0.05$	&	---	&	4.18$_{-0.12}^{+0.11}$	&	---	&	---	&	---\\
NGC\,6210	&	5.74$_{-0.16}^{+0.11}$	&	---	&	4.08$_{-0.11}^{+0.08}$	&	---	&	---	&	---\\
NGC\,6302	&	---	&	5.54$_{-0.07}^{+0.06}$	&	4.21$_{-0.04}^{+0.03}$	&	4.10$\pm0.05$	&	4.00$\pm0.04$	&	---\\
NGC\,6302	&	---	&	---	&	4.35$_{-0.29}^{+0.17}$	&	---	&	3.97$_{-0.28}^{+0.16}$	&	4.94$_{-0.26}^{+0.15}$\\
NGC\,6309	&	---	&	---	&	4.45$_{-0.31}^{+0.19}$	&	---	&	---	&	---\\
NGC\,6369	&	---	&	---	&	3.85$\pm0.08$	&	---	&	---	&	---\\
NGC\,6537	&	---	&	---	&	4.43$_{-0.10}^{+0.12}$	&	---	&	4.17$_{-0.12}^{+0.13}$	&	4.77$_{-0.18}^{+0.12}$\\
NGC\,6543	&	---	&	---	&	3.67$_{-0.15}^{+0.10}$	&	---	&	---	&	---\\
NGC\,6572	&	---	&	---	&	3.91$_{-0.15}^{+0.11}$	&	---	&	---	&	---\\
NGC\,6790	&	---	&	---	&	3.68$_{-0.14}^{+0.12}$	&	---	&	---	&	---\\
NGC\,6818	&	---	&	---	&	4.17$_{-0.04}^{+0.03}$	&	---	&	---	&	---\\
NGC\,6884	&	---	&	---	&	4.36$_{-0.28}^{+0.19}$	&	---	&	---	&	4.95$_{-0.18}^{+0.12}$\\
NGC\,6886	&	---	&	6.12$_{-0.11}^{+0.09}$	&	4.26$_{-0.13}^{+0.12}$	&	---	&	---	&	4.78$_{-0.10}^{+0.09}$\\
NGC\,6886	&	---	&	---	&	4.03$_{-0.29}^{+0.16}$	&	---	&	---	&	4.77$_{-0.20}^{+0.16}$\\
NGC\,7026	&	---	&	---	&	4.62$_{-0.27}^{+0.18}$	&	---	&	---	&	---\\
M\,1-33	&	---	&	---	&	3.91$_{-0.14}^{+0.13}$	&	---	&	---	&	---\\
M\,1-50	&	---	&	---	&	4.18$_{-0.22}^{+0.16}$	&	---	&	---	&	---\\
M\,1-54	&	---	&	---	&	3.92$_{-0.24}^{+0.16}$	&	---	&	---	&	---\\
M\,1-57	&	---	&	---	&	4.58$_{-0.27}^{+0.15}$	&	---	&	---	&	6.28$_{-0.28}^{+0.15}$\\
M\,1-60	&	---	&	---	&	4.07$_{-0.11}^{+0.12}$	&	---	&	---	&	---\\
M\,1-61	&	---	&	---	&	3.41$_{-0.11}^{+0.09}$	&	---	&	---	&	---\\
M\,1-80	&	---	&	---	&	---	&	---	&	---	&	5.06$_{-0.30}^{+0.20}$\\
M\,2-31	&	---	&	---	&	3.99$\pm0.09$	&	---	&	---	&	---\\
PC\,14	&	---	&	---	&	3.99$\pm0.07$	&	---	&	---	&	---\\
\hline
\end{tabular}
\end{threeparttable}
\label{tab:5}
\end{table*}

\begin{table*}
\centering
\caption{Ionic abundances of oxygen, helium and argon in 12$+\log$(X$^{+i}$/H$^{+}$).}
\begin{threeparttable}
\begin{tabular}{lcccccccc}
\hline         
Name                    & O$^{+}$/H$^{+}$                        & O$^{++}$/H$^{+}$                  & He$^{+}$/H$^{+}$                        & He$^{++}$/H$^{+}$                          & Ar$^{++}$/H$^{+}$ & Ar$^{+3}$/H$^{+}$ & Ar$^{+4}$/H$^{+}$\\  
\hline     
Fg\,1	&	6.59$\pm0.06$	&	8.24$_{-0.05}^{+0.04}$	&	11.19$_{-0.04}^{+0.05}$	&	10.45$_{-0.04}^{+0.03}$	&	5.96$_{-0.11}^{+0.08}$	&	5.92$_{-0.05}^{+0.04}$	&	---\\
H\,1-40	&	7.11$_{-0.18}^{+0.36}$	&	8.50$_{-0.06}^{+0.07}$	&	11.10$\pm0.02$	&	---	&	6.17$_{-0.05}^{+0.04}$	&	4.80$_{-0.23}^{+0.19}$	&	---\\
H\,1-50	&	7.31$_{-0.10}^{+0.13}$	&	8.59$\pm0.04$	&	10.98$_{-0.02}^{+0.03}$	&	10.03$\pm0.02$	&	5.86$_{-0.06}^{+0.10}$	&	5.88$_{-0.04}^{+0.05}$	&	4.87$\pm0.04$\\
Hb\,4	&	7.35$_{-0.18}^{+0.34}$	&	8.35$_{-0.06}^{+0.05}$	&	10.98$_{-0.04}^{+0.03}$	&	10.31$_{-0.03}^{+0.02}$	&	6.25$\pm0.07$	&	6.05$_{-0.04}^{+0.07}$	&	4.35$\pm0.06$\\
Hb\,5	&	7.65$_{-0.06}^{+0.08}$	&	8.47$_{-0.07}^{+0.05}$	&	10.81$\pm0.04$	&	10.72$\pm0.03$	&	6.44$\pm0.04$	&	6.29$\pm0.06$	&	5.86$\pm0.05$\\
He\,2-73	&	7.63$_{-0.10}^{+0.15}$	&	8.49$_{-0.05}^{+0.04}$	&	10.96$_{-0.03}^{+0.02}$	&	10.30$\pm0.02$	&	6.15$_{-0.07}^{+0.06}$	&	5.98$\pm0.06$	&	5.17$_{-0.04}^{+0.03}$\\
Hen\,2-86	&	7.38$_{-0.15}^{+0.21}$	&	8.74$\pm0.05$	&	11.12$\pm0.03$	&	---	&	6.56$\pm0.05$	&	5.57$_{-0.04}^{+0.06}$	&	---\\
Hu\,1-2	&	7.11$_{-0.13}^{+0.16}$	&	7.56$_{-0.06}^{+0.07}$	&	10.60$\pm0.04$	&	10.83$\pm0.03$	&	5.38$\pm0.05$	&	5.39$\pm0.06$	&	5.03$_{-0.07}^{+0.06}$\\
IC\,2165	&	7.10$_{-0.06}^{+0.05}$	&	8.21$_{-0.02}^{+0.03}$	&	10.84$_{-0.03}^{+0.02}$	&	10.57$\pm0.01$	&	5.81$\pm0.03$	&	5.73$\pm0.02$	&	5.16$_{-0.12}^{+0.10}$\\
IC\,2165	&	6.86$_{-0.21}^{+0.31}$	&	8.13$\pm0.06$	&	10.63$\pm0.05$	&	10.74$_{-0.03}^{+0.02}$	&	5.48$_{-0.07}^{+0.08}$	&	5.76$_{-0.06}^{+0.05}$	&	5.36$\pm0.04$\\
IC\,4191	&	7.42$\pm0.08$	&	8.73$_{-0.04}^{+0.03}$	&	11.04$\pm0.02$	&	10.04$\pm0.02$	&	6.36$\pm0.04$	&	6.10$_{-0.04}^{+0.03}$	&	5.19$\pm0.05$\\
IC\,5217	&	6.70$_{-0.22}^{+0.25}$	&	8.63$_{-0.05}^{+0.06}$	&	10.89$_{-0.04}^{+0.03}$	&	9.96$\pm0.03$	&	5.93$\pm0.09$	&	5.99$\pm0.06$	&	4.32$_{-0.17}^{+0.14}$\\
IC\,5217	&	6.78$_{-0.20}^{+0.36}$	&	8.47$\pm0.09$	&	10.96$_{-0.04}^{+0.05}$	&	9.90$\pm0.04$	&	5.89$_{-0.05}^{+0.06}$	&	5.91$_{-0.07}^{+0.09}$	&	4.04$_{-0.29}^{+0.17}$\\
NGC\,2022	&	6.17$_{-0.06}^{+0.05}$	&	7.90$_{-0.05}^{+0.04}$	&	10.10$_{-0.02}^{+0.03}$	&	10.99$\pm0.02$	&	5.54$\pm0.05$	&	6.08$_{-0.04}^{+0.03}$	&	5.50$\pm0.04$\\
NGC\,2867-1	&	7.40$_{-0.07}^{+0.08}$	&	8.36$\pm0.04$	&	10.83$\pm0.02$	&	10.51$\pm0.02$	&	5.86$\pm0.05$	&	5.62$_{-0.04}^{+0.05}$	&	4.52$_{-0.05}^{+0.04}$\\
NGC\,2867-2	&	7.59$\pm0.08$	&	8.45$\pm0.04$	&	10.92$\pm0.02$	&	10.46$\pm0.02$	&	6.02$\pm0.06$	&	5.55$\pm0.04$	&	4.47$_{-0.07}^{+0.06}$\\
NGC\,3242	&	6.73$_{-0.16}^{+0.19}$	&	8.43$_{-0.05}^{+0.04}$	&	10.89$\pm0.02$	&	10.34$\pm0.02$	&	5.80$_{-0.04}^{+0.05}$	&	5.92$\pm0.04$	&	5.74$_{-0.10}^{+0.08}$\\
NGC\,3918	&	7.61$_{-0.10}^{+0.11}$	&	8.41$\pm0.04$	&	10.76$\pm0.03$	&	10.61$\pm0.02$	&	6.01$_{-0.05}^{+0.04}$	&	6.05$\pm0.04$	&	5.39$_{-0.04}^{+0.03}$\\
NGC\,5189	&	8.13$_{-0.09}^{+0.08}$	&	8.41$\pm0.04$	&	10.92$\pm0.03$	&	10.61$\pm0.02$	&	6.37$_{-0.04}^{+0.05}$	&	5.83$\pm0.04$	&	4.82$\pm0.05$\\
NGC\,5882	&	6.91$_{-0.07}^{+0.08}$	&	8.66$\pm0.04$	&	11.02$\pm0.02$	&	9.35$\pm0.02$	&	6.15$\pm0.03$	&	6.03$_{-0.03}^{+0.04}$	&	---\\
NGC\,6153	&	7.21$_{-0.20}^{+0.25}$	&	8.64$\pm0.06$	&	11.10$\pm0.03$	&	10.02$\pm0.03$	&	6.37$_{-0.09}^{+0.08}$	&	6.03$_{-0.06}^{+0.07}$	&	---\\
NGC\,6210	&	7.25$_{-0.14}^{+0.16}$	&	8.63$\pm0.09$	&	10.96$_{-0.05}^{+0.04}$	&	9.09$\pm0.04$	&	6.00$_{-0.06}^{+0.07}$	&	5.68$_{-0.08}^{+0.07}$	&	---\\
NGC\,6302	&	7.23$_{-0.10}^{+0.11}$	&	7.93$_{-0.05}^{+0.04}$	&	10.86$_{-0.03}^{+0.02}$	&	10.83$\pm0.02$	&	6.07$\pm0.05$	&	6.00$\pm0.04$	&	5.75$_{-0.05}^{+0.03}$\\
NGC\,6302	&	6.92$_{-0.25}^{+0.62}$	&	8.14$_{-0.05}^{+0.07}$	&	10.83$\pm0.03$	&	10.83$\pm0.02$	&	6.05$_{-0.15}^{+0.16}$	&	6.21$_{-0.04}^{+0.05}$	&	5.85$\pm0.05$\\
NGC\,6309	&	7.01$_{-0.45}^{+0.64}$	&	8.31$_{-0.08}^{+0.09}$	&	10.66$_{-0.06}^{+0.05}$	&	10.86$_{-0.05}^{+0.04}$	&	5.88$\pm0.06$	&	6.12$\pm0.08$	&	5.57$\pm0.06$\\
NGC\,6369	&	6.94$_{-0.08}^{+0.09}$	&	8.51$_{-0.04}^{+0.05}$	&	11.03$_{-0.03}^{+0.02}$	&	8.60$_{-0.03}^{+0.04}$	&	6.10$\pm0.06$	&	5.45$\pm0.04$	&	---\\
NGC\,6537	&	6.82$_{-0.13}^{+0.14}$	&	7.91$\pm0.13$	&	10.76$\pm0.05$	&	11.00$_{-0.10}^{+0.09}$	&	5.90$_{-0.13}^{+0.14}$	&	6.12$_{-0.13}^{+0.14}$	&	6.10$\pm0.10$\\
NGC\,6543	&	7.29$_{-0.15}^{+0.23}$	&	8.76$\pm0.05$	&	11.07$\pm0.02$	&	---	&	6.49$\pm0.03$	&	5.77$\pm0.05$	&	---\\
NGC\,6572	&	6.84$_{-0.27}^{+0.42}$	&	8.60$\pm0.09$	&	11.08$_{-0.05}^{+0.04}$	&	8.69$_{-0.11}^{+0.09}$	&	6.20$\pm0.06$	&	5.74$\pm0.09$	&	3.88$_{-0.31}^{+0.18}$\\
NGC\,6790	&	6.15$_{-0.33}^{+0.35}$	&	8.39$_{-0.10}^{+0.07}$	&	11.08$_{-0.06}^{+0.05}$	&	9.50$_{-0.05}^{+0.04}$	&	5.65$_{-0.06}^{+0.05}$	&	5.58$_{-0.09}^{+0.08}$	&	4.01$_{-0.29}^{+0.19}$\\
NGC\,6818	&	7.38$_{-0.06}^{+0.05}$	&	8.35$_{-0.04}^{+0.03}$	&	10.67$\pm0.02$	&	10.72$\pm0.02$	&	6.00$_{-0.04}^{+0.05}$	&	5.96$\pm0.04$	&	5.19$\pm0.06$\\
NGC\,6884	&	7.11$_{-0.31}^{+0.41}$	&	8.61$_{-0.08}^{+0.09}$	&	10.98$_{-0.04}^{+0.05}$	&	10.23$_{-0.05}^{+0.04}$	&	6.14$\pm0.05$	&	6.09$_{-0.07}^{+0.08}$	&	4.66$_{-0.20}^{+0.15}$\\
NGC\,6886	&	7.94$_{-0.16}^{+0.35}$	&	8.57$\pm0.06$	&	10.75$_{-0.05}^{+0.04}$	&	10.59$_{-0.03}^{+0.02}$	&	6.27$\pm0.09$	&	6.13$_{-0.07}^{+0.08}$	&	5.50$_{-0.04}^{+0.05}$\\
NGC\,6886	&	7.91$_{-0.26}^{+0.44}$	&	8.44$_{-0.10}^{+0.09}$	&	10.89$\pm0.05$	&	10.53$_{-0.05}^{+0.04}$	&	6.18$\pm0.06$	&	5.98$_{-0.08}^{+0.07}$	&	5.40$\pm0.09$\\
NGC\,7026	&	7.59$_{-0.17}^{+0.23}$	&	8.74$_{-0.11}^{+0.15}$	&	11.07$\pm0.05$	&	10.11$_{-0.04}^{+0.05}$	&	6.47$_{-0.08}^{+0.10}$	&	6.15$_{-0.12}^{+0.15}$	&	---\\
M\,1-33	&	7.76$\pm0.10$	&	8.72$\pm0.04$	&	11.11$\pm0.02$	&	9.15$\pm0.03$	&	6.50$\pm0.07$	&	5.83$\pm0.05$	&	---\\
M\,1-50	&	7.12$_{-0.19}^{+0.33}$	&	8.65$_{-0.09}^{+0.10}$	&	10.94$\pm0.05$	&	10.30$\pm0.04$	&	5.99$\pm0.07$	&	6.05$_{-0.08}^{+0.09}$	&	4.23$_{-0.14}^{+0.10}$\\
M\,1-54	&	8.47$_{-0.25}^{+0.42}$	&	8.47$_{-0.09}^{+0.10}$	&	11.10$_{-0.05}^{+0.04}$	&	10.20$_{-0.05}^{+0.04}$	&	6.36$_{-0.06}^{+0.05}$	&	5.77$_{-0.09}^{+0.08}$	&	4.90$_{-0.12}^{+0.08}$\\
M\,1-57	&	7.66$_{-0.20}^{+0.24}$	&	8.50$_{-0.08}^{+0.09}$	&	10.89$\pm0.05$	&	10.62$_{-0.05}^{+0.04}$	&	6.24$\pm0.06$	&	6.04$\pm0.08$	&	5.73$_{-0.06}^{+0.07}$\\
M\,1-60	&	7.79$_{-0.10}^{+0.14}$	&	8.76$\pm0.05$	&	11.12$\pm0.02$	&	8.94$_{-0.04}^{+0.03}$	&	6.56$\pm0.10$	&	5.88$\pm0.05$	&	---\\
M\,1-61	&	7.21$_{-0.16}^{+0.19}$	&	8.64$_{-0.04}^{+0.05}$	&	11.06$\pm0.03$	&	---	&	6.40$\pm0.05$	&	5.40$_{-0.04}^{+0.06}$	&	---\\
M\,1-80	&	7.77$_{-0.11}^{+0.14}$	&	8.63$_{-0.09}^{+0.08}$	&	10.82$_{-0.05}^{+0.03}$	&	10.48$_{-0.05}^{+0.04}$	&	5.99$_{-0.05}^{+0.06}$	&	5.71$_{-0.08}^{+0.07}$	&	4.83$_{-0.25}^{+0.16}$\\
M\,2-31	&	7.50$_{-0.09}^{+0.11}$	&	8.59$\pm0.05$	&	11.05$\pm0.02$	&	8.56$\pm0.06$	&	6.15$_{-0.11}^{+0.09}$	&	5.75$_{-0.04}^{+0.05}$	&	---\\
PC\,14	&	7.37$_{-0.09}^{+0.08}$	&	8.73$_{-0.05}^{+0.04}$	&	11.02$\pm0.03$	&	9.58$\pm0.02$	&	6.27$\pm0.06$	&	5.71$\pm0.04$	&	---\\
\hline
\end{tabular}
\end{threeparttable}
\label{tab:6}
\end{table*}

\begin{table*}
\caption{Total abundances in 12$+\log$(X/H).}
\begin{threeparttable}
\begin{tabular}{lcccccll}
\hline   
Name    & O/H   & Ar/H   & Na/H   & K/H   &  Ca/H    & O$^{++}$/(O$^{+}$+O$^{++}$)   & He$^{++}$/(He$^{+}$+He$^{++}$)\\
\hline  
Fg\,1	&	8.62$_{-0.08}^{+0.07}$	&	6.63$_{-0.41}^{+0.20}$	&	---	&	5.24$_{-0.40}^{+0.27}$	&	---	&	0.979$_{-0.003}^{+0.001}$	&	0.15$_{-0.04}^{+0.02}$\\
H\,1-40	&	8.65$\pm0.09$	&	6.54$_{-0.38}^{+0.39}$	&	---	&	---	&	---	&	0.96$_{-0.04}^{+0.01}$	&	0.00					\\
H\,1-50	&	8.35$_{-0.05}^{+0.06}$	&	6.27$_{-0.32}^{+0.36}$	&	---	&	4.90$_{-0.36}^{+0.40}$	&	---	&	0.95$_{-0.02}^{+0.01}$	&	0.10$\pm0.01$\\
Hb\,4	&	8.40$\pm0.09$	&	6.45$_{-0.33}^{+0.19}$	&	---	&	4.83$_{-0.33}^{+0.21}$	&	---	&	0.91$_{-0.08}^{+0.03}$	&	0.18$\pm0.01$\\
Hb\,5	&	8.51$\pm0.09$	&	6.73$_{-0.26}^{+0.15}$	&	6.46$\pm0.07$	&	---	&	---	&	0.874$_{-0.018}^{+0.003}$	&	0.45$_{-0.03}^{+0.02}$\\
He\,2-73	&	8.66$_{-0.05}^{+0.06}$	&	6.34$_{-0.26}^{+0.17}$	&	---	&	4.58$_{-0.26}^{+0.19}$	&	5.48$_{-0.28}^{+0.24}$	&	0.89$_{-0.05}^{+0.02}$	&	0.18$\pm0.01$\\
Hen\,2-86	&	8.72$\pm0.10$	&	7.00$_{-0.41}^{+0.31}$	&	---	&	---	&	---	&	0.958$_{-0.024}^{+0.012}$	&	0.00					\\
Hu\,1-2	&	8.81$_{-0.12}^{+0.26}$	&	5.74$_{-0.29}^{+0.17}$	&	---	&	4.35$_{-0.28}^{+0.22}$	&	---	&	0.76$_{-0.11}^{+0.03}$	&	0.62$_{-0.02}^{+0.03}$\\
IC\,2165	&	8.68$_{-0.09}^{+0.10}$	&	6.08$_{-0.27}^{+0.17}$	&	---	&	4.56$_{-0.27}^{+0.18}$	&	4.67$_{-0.36}^{+0.29}$	&	0.93$\pm0.01$	&	0.35$\pm0.01$\\
IC\,2165	&	8.79$_{-0.08}^{+0.11}$	&	6.04$_{-0.37}^{+0.40}$	&	5.51$_{-0.21}^{+0.30}$	&	4.51$_{-0.41}^{+0.44}$	&	4.74$_{-0.42}^{+0.41}$	&	0.95$_{-0.04}^{+0.02}$	&	0.56$\pm0.03$\\
IC\,4191	&	8.64$_{-0.12}^{+0.15}$	&	6.80$_{-0.37}^{+0.35}$	&	---	&	5.36$_{-0.42}^{+0.32}$	&	---	&	0.95$\pm0.01$	&	0.09$\pm0.01$\\
IC\,5217	&	8.61$_{-0.09}^{+0.10}$	&	6.56$_{-0.39}^{+0.23}$	&	6.56$_{-0.22}^{+0.25}$	&	5.02$_{-0.42}^{+0.22}$	&	---	&	0.990$_{-0.012}^{+0.003}$	&	0.10$\pm0.01$\\
IC\,5217	&	8.40$_{-0.10}^{+0.07}$	&	6.50$_{-0.51}^{+0.22}$	&	---	&	4.55$_{-0.43}^{+0.36}$	&	---	&	0.981$_{-0.025}^{+0.006}$	&	0.08$\pm0.01$\\
NGC\,2022	&	8.67$_{-0.10}^{+0.09}$	&	6.87$_{-0.39}^{+0.25}$	&	---	&	5.48$_{-0.38}^{+0.25}$	&	---	&	0.982$_{-0.002}^{+0.001}$	&	0.89$\pm0.01$\\
NGC\,2867-1	&	8.81$_{-0.08}^{+0.12}$	&	6.09$_{-0.29}^{+0.19}$	&	---	&	4.65$_{-0.27}^{+0.22}$	&	---	&	0.90$_{-0.02}^{+0.01}$	&	0.32$_{-0.01}^{+0.02}$\\
NGC\,2867-2	&	8.67$_{-0.13}^{+0.16}$	&	6.19$_{-0.31}^{+0.20}$	&	---	&	4.96$_{-0.31}^{+0.20}$	&	---	&	0.88$\pm0.02$	&	0.26$_{-0.01}^{+0.02}$\\
NGC\,3242	&	8.81$_{-0.12}^{+0.17}$	&	6.43$_{-0.43}^{+0.24}$	&	---	&	4.91$_{-0.44}^{+0.24}$	&	---	&	0.98$\pm0.01$	&	0.22$\pm0.01$\\
NGC\,3918	&	8.66$_{-0.12}^{+0.08}$	&	6.26$_{-0.28}^{+0.17}$	&	6.12$_{-0.22}^{+0.21}$	&	4.70$_{-0.26}^{+0.19}$	&	4.97$_{-0.29}^{+0.18}$	&	0.87$_{-0.04}^{+0.02}$	&	0.42$\pm0.02$\\
NGC\,5189	&	8.50$_{-0.05}^{+0.04}$	&	6.49$_{-0.28}^{+0.16}$	&	---	&	4.62$_{-0.27}^{+0.20}$	&	---	&	0.66$\pm0.05$	&	0.33$\pm0.02$\\
NGC\,5882	&	8.68$\pm0.04$	&	6.72$_{-0.38}^{+0.24}$	&	---	&	5.08$_{-0.37}^{+0.29}$	&	---	&	0.983$_{-0.004}^{+0.002}$	&	0.021$_{-0.001}^{+0.002}$\\
NGC\,6153	&	8.20$\pm0.07$	&	6.86$_{-0.35}^{+0.32}$	&	6.37$_{-0.04}^{+0.05}$	&	5.27$_{-0.41}^{+0.33}$	&	---	&	0.97$_{-0.03}^{+0.01}$	&	0.08$\pm0.01$\\
NGC\,6210	&	8.38$\pm0.11$	&	6.45$_{-0.43}^{+0.32}$	&	5.81$_{-0.12}^{+0.07}$	&	5.22$_{-0.56}^{+0.27}$	&	---	&	0.96$_{-0.02}^{+0.01}$	&	0.013$\pm0.002$\\
NGC\,6302	&	8.61$\pm0.08$	&	6.35$_{-0.25}^{+0.18}$	&	5.85$_{-0.22}^{+0.20}$	&	4.75$_{-0.26}^{+0.18}$	&	---	&	0.84$_{-0.05}^{+0.02}$	&	0.48$\pm0.02$\\
NGC\,6302	&	8.77$\pm0.04$	&	6.50$_{-0.35}^{+0.37}$	&	---	&	4.79$_{-0.44}^{+0.41}$	&	5.66$_{-0.45}^{+0.51}$	&	0.91$_{-0.09}^{+0.06}$	&	0.50$\pm0.02$\\
NGC\,6309	&	8.78$\pm0.05$	&	6.49$_{-0.41}^{+0.45}$	&	---	&	4.98$_{-0.42}^{+0.48}$	&	---	&	0.96$_{-0.14}^{+0.02}$	&	0.61$_{-0.08}^{+0.04}$\\
NGC\,6369	&	8.77$_{-0.04}^{+0.06}$	&	6.75$_{-0.39}^{+0.16}$	&	---	&	---	&	---	&	0.974$_{-0.007}^{+0.005}$	&	0.0037$_{-0.0004}^{+0.0005}$\\
NGC\,6537	&	8.78$_{-0.07}^{+0.11}$	&	6.41$_{-0.30}^{+0.35}$	&	---	&	4.93$_{-0.30}^{+0.35}$	&	5.20$_{-0.33}^{+0.35}$	&	0.926$_{-0.041}^{+0.023}$	&	0.63$\pm0.06$\\
NGC\,6543	&	8.55$_{-0.08}^{+0.12}$	&	7.05$_{-0.44}^{+0.25}$	&	---	&	---	&	---	&	0.97$_{-0.02}^{+0.01}$	&	0.00					\\
NGC\,6572	&	8.64$_{-0.07}^{+0.14}$	&	6.75$_{-0.45}^{+0.25}$	&	---	&	---	&	---	&	0.981$_{-0.025}^{+0.009}$	&	0.004$\pm0.001$\\
NGC\,6790	&	8.45$_{-0.07}^{+0.08}$	&	6.28$_{-0.39}^{+0.22}$	&	---	&	4.65$_{-0.35}^{+0.30}$	&	---	&	0.9968$_{-0.0096}^{+0.0004}$	&	0.02$_{-0.02}^{+0.01}$\\
NGC\,6818	&	8.66$\pm0.05$	&	6.31$_{-0.23}^{+0.19}$	&	---	&	4.71$_{-0.24}^{+0.18}$	&	---	&	0.90$_{-0.02}^{+0.01}$	&	0.53$\pm0.02$\\
NGC\,6884	&	8.70$_{-0.06}^{+0.07}$	&	6.69$_{-0.44}^{+0.31}$	&	---	&	5.11$_{-0.43}^{+0.37}$	&	7.11$_{-0.46}^{+0.41}$	&	0.971$_{-0.043}^{+0.012}$	&	0.15$\pm0.02$\\
NGC\,6886	&	8.53$\pm0.04$	&	6.49$_{-0.28}^{+0.16}$	&	6.45$_{-0.23}^{+0.22}$	&	4.72$_{-0.23}^{+0.26}$	&	5.84$_{-0.27}^{+0.25}$	&	0.80$_{-0.12}^{+0.07}$	&	0.41$\pm0.03$\\
NGC\,6886	&	8.76$\pm0.04$	&	6.32$_{-0.25}^{+0.20}$	&	---	&	4.49$_{-0.34}^{+0.32}$	&	5.87$_{-0.31}^{+0.21}$	&	0.76$_{-0.21}^{+0.10}$	&	0.30$\pm0.03$\\
NGC\,7026	&	8.53$_{-0.06}^{+0.08}$	&	6.74$_{-0.31}^{+0.46}$	&	---	&	5.47$\pm0.42$	&	---	&	0.94$_{-0.05}^{+0.02}$	&	0.10$\pm0.01$\\
M\,1-33	&	8.77$_{-0.05}^{+0.04}$	&	6.65$_{-0.29}^{+0.16}$	&	---	&	---	&	---	&	0.91$_{-0.03}^{+0.01}$	&	0.011$\pm0.001$\\
M\,1-50	&	8.81$\pm0.05$	&	6.55$_{-0.41}^{+0.30}$	&	---	&	4.82$_{-0.40}^{+0.37}$	&	---	&	0.97$_{-0.03}^{+0.01}$	&	0.19$_{-0.03}^{+0.02}$\\
M\,1-54	&	8.63$\pm0.04$	&	6.44$_{-0.29}^{+0.17}$	&	---	&	4.79$_{-0.35}^{+0.26}$	&	---	&	0.48$_{-0.20}^{+0.18}$	&	0.11$_{-0.02}^{+0.01}$\\
M\,1-57	&	8.29$\pm0.05$	&	6.47$_{-0.26}^{+0.19}$	&	---	&	5.14$_{-0.34}^{+0.25}$	&	7.12$_{-0.37}^{+0.23}$	&	0.89$_{-0.09}^{+0.03}$	&	0.35$\pm0.03$\\
M\,1-60	&	8.01$_{-0.09}^{+0.10}$	&	6.67$_{-0.28}^{+0.22}$	&	---	&	---	&	---	&	0.91$_{-0.04}^{+0.01}$	&	0.006$\pm0.001$\\
M\,1-61	&	8.27$_{-0.14}^{+0.13}$	&	6.95$_{-0.44}^{+0.25}$	&	---	&	---	&	---	&	0.940$_{-0.001}^{+0.034}$	&	0.00					\\
M\,1-80	&	8.51$_{-0.07}^{+0.06}$	&	6.18$_{-0.26}^{+0.21}$	&	---	&	---	&	6.54$_{-0.46}^{+0.37}$	&	0.88$_{-0.05}^{+0.03}$	&	0.31$_{-0.02}^{+0.04}$\\
M\,2-31	&	8.59$_{-0.06}^{+0.05}$	&	6.31$_{-0.28}^{+0.22}$	&	---	&	---	&	---	&	0.93$_{-0.03}^{+0.01}$	&	0.0032$\pm0.0005$\\
PC\,14	&	8.68$_{-0.06}^{+0.07}$	&	6.79$_{-0.44}^{+0.28}$	&	---	&	5.38$_{-0.45}^{+0.27}$	&	---	&	0.96$\pm0.01$	&	0.035$_{-0.002}^{+0.003}$\\
\hline
\end{tabular}
\end{threeparttable}
\label{tab:7}
\end{table*}

\end{document}